\documentclass[1p]{elsarticle}

\usepackage{placeins}
\usepackage{lineno,hyperref}
\modulolinenumbers[5]
\usepackage{amsmath}
\usepackage{siunitx}

\newcommand{\WfourtyIF}{W40\textunderscore IF}
\newcommand{\WelevenOFtwo}{W11\textunderscore OF2}
\newcommand{\WfiveOBone}{W05\textunderscore OB1}
\newcommand{\WfiveOBoneIrrad}{W05\textunderscore OB1*}
\newcommand{\WphaseTwo}{W37\textunderscore OF2}

\journal{Nuclear Inst. and Methods in Physics Research, A}

\begin{document}

\begin{frontmatter}

\title{Performance of production modules of the Belle~II pixel detector in a high-energy particle beam}

\author[UGoett]{P.~Wieduwilt\corref{cor1}}
\ead{philipp.wieduwilt@phys.uni-goettingen.de}
\cortext[cor1]{Corresponding author}
\author[UBonn]{B.~Paschen\corref{cor1}}
\ead{paschen@physik.uni-bonn.de}
\author[UGoett]{H.~Schreeck}
\author[UGoett]{B.~Schwenker}
\author[UGoett]{J.~Soltau}
\author[UBonn]{P.~Ahlburg}
\author[UBonn]{J.~Dingfelder}
\author[UGoett]{A.~Frey}
\author[IFIC]{P.~Gomis}
\author[UBonn]{F.~L\"utticke}
\author[IFIC]{C.~Marinas} 

\address[UBonn]{Physikalisches Institut, Universit\"at Bonn, Nu{\ss}allee 12, 53115 Bonn, Germany}
\address[UGoett]{II. Physikalisches Institut, Georg-August-Universit\"at G\"ottingen, Friedrich-Hund-Platz 1, 37077 G\"ottingen, Germany}
\address[IFIC]{IFIC (UVEG/CSIC), Catedr\'atico, Jos\'e Beltr\'an 2, E-46071 Valencia, Spain}

\begin{abstract}
The Belle~II experiment at the Super B factory SuperKEKB, an asymmetric $e^+e^-$ collider located in Tsukuba, Japan, is tailored to perform precision B physics measurements. The centre of mass energy of the collisions is equal to the rest mass of the $\Upsilon(4S)$ resonance of $m_{\Upsilon(4S)} = \SI{10.58}{\giga\electronvolt}$. A high vertex resolution is essential for measuring the decay vertices of B mesons. Typical momenta of the decay products are ranging from a few tens of \si{\mega\electronvolt} to a few \si{\giga\electronvolt} and multiple scattering has a significant impact on the vertex resolution. The VerteX Detector (VXD) for Belle~II is therefore designed to have as little material as possible inside the acceptance region. Especially the innermost two layers, populated by the PiXel Detector (PXD), have to be ultra-thin. The PXD is based on \textit{DEpleted P-channel Field Effect Transistors} (DEPFETs) with a thickness of only \SI{75}{\micro\metre}. Spatial resolution and hit efficiency of production detector modules were studied in beam tests performed at the DESY test beam facility. The spatial resolution was investigated as a function of the incidence angle and improvements due to charge sharing are demonstrated. The measured module performance is compatible with the requirements for Belle~II. 
\end{abstract}


\end{frontmatter}

\section{Introduction}
The Belle~II experiment~\cite{Abe:2010gxa} started taking data in spring 2019 at the SuperKEKB~\cite{ohnishi2013accelerator} $e^+e^-$ collider in Tsukuba, Japan.
SuperKEKB aims at a target luminosity of \SI{8E35}{\per\centi\metre\squared\per\second}.
The experiment consists of a detector around the interaction point where electrons and positrons collide with a centre of mass energy of $E_\mathrm{CM} = m_{\Upsilon(4S)} = \SI{10.58}{\GeV}$ corresponding to the mass of the $\Upsilon(4S)$ resonance.
The detector is composed of several subsystems for tracking, particle identification and calorimetry.
It covers a large solid angle with cylindrical geometry and a polar angle tracking acceptance from $\SI{17}{\degree}$ to $\SI{150}{\degree}$.
The central VerteX Detector (VXD) features six layers of silicon detectors~\cite{Adachi2018}.
Closest to the interaction point is the PiXel Detector (PXD) arranged in two cylindrical layers at radii of \SI{14}{mm} and \SI{22}{mm} from the beam axis.
Its final design consists of 40 modules, which are glued end-to-end in pairs of two, forming eight inner and twelve outer ladders.
The PXD sensors are based on DEpleted P-channel Field Effect Transistors (DEPFET)~\cite{KEMMER1987365}.
DEPFET is a fully depleted active pixel sensor technology, which is used for the first time in a high-energy physics experiment.

In the current detector, a reduced PXD with 19 functional modules was installed. They have been reported to yield hit efficiencies above \SI{98}{\percent} ~\cite{Spruck_Vertex2019, Schwenker_Vertex2018}.

Due to a reduced boost of the collision centre of mass system, a better vertex resolution is needed in Belle~II compared to its predecessor experiment.
From simulations with final design parameters, the VXD is expected to yield a two times better impact parameter resolution compared to the Belle vertex detector of $\sim\SI{10}{\micro\metre}$ for high momentum ($p>\SI{1}{\giga\electronvolt}$) particles \cite{Kou_2019}.
Preliminary measurements determined an impact parameter resolution of \SI{14.1\pm0.1}{\micro\metre} \cite{Bilka_Vertex2019}.

The results of the first dedicated beam test of final production modules are presented in the following.
The measurements were carried out at the DESY beam test facility and facilitate efficiency studies over the whole sensor region and with sub-pixel resolution.
To evaluate the effect of radiation damage, one of the modules was tested before and after X-ray irradiation with the estimated Belle~II PXD lifetime dose of $\sim\SI{200}{\kilo\gray}$~\cite{Schreeck2020} on the sensor.
The module design, its operation and the beam test setup are introduced.
This is followed by a description of the reconstruction and analysis procedures.
Finally, resolution and efficiency studies from beam test data are presented.
For the resolution study, a new Position Finding Algorithm (PFA) based on cluster shapes is introduced and compared with the conventional Centre of Gravity (CoG) approach.
This cluster shape PFA could be used for Belle~II as an alternative to the CoG PFA approach.

\section{DEPFET modules}
\begin{figure}[p]
  \centering
  \includegraphics[clip,trim=0cm 4.5cm 0cm 2.5cm,width=\textwidth]{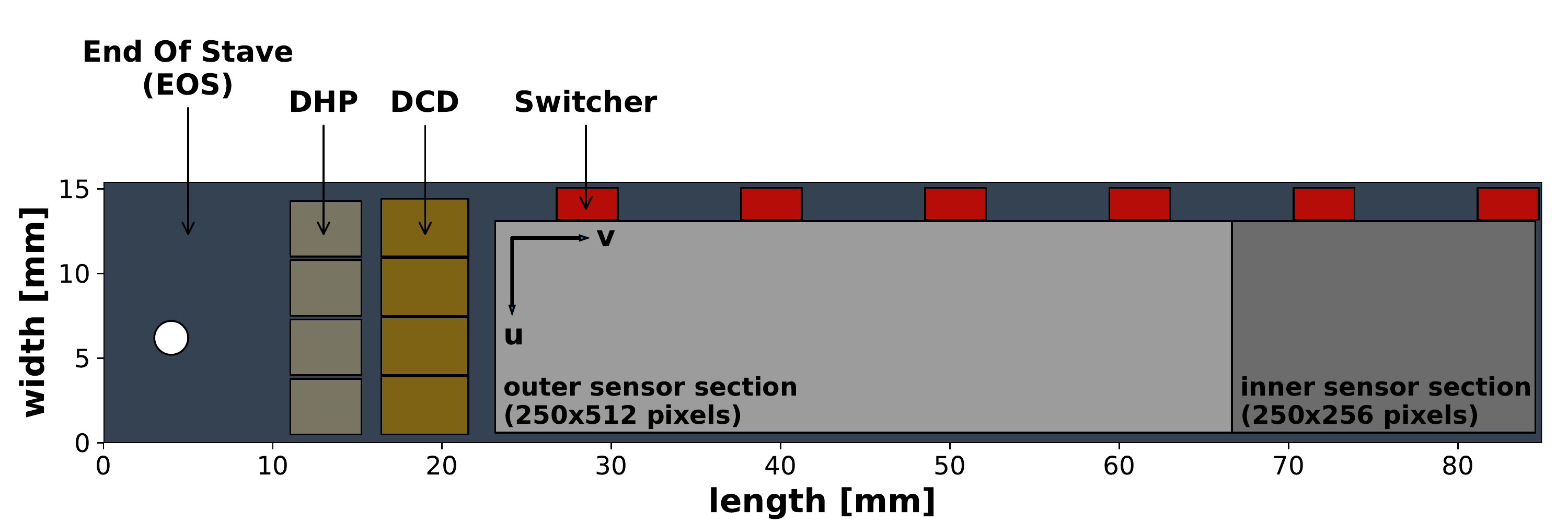}
  \caption{Schematic drawing of the layout of an outer module for the pixel detector. The read-out ASICs (DHP, DCD, Switcher) are indicated, as well as the End of Stave (EOS) region.}
  \label{fig:module_layout}
\end{figure}

% Module introduction
Each module in the first (second) layer is made up of a large $1.5\times6.8(8.5)~\si{\centi\metre\squared}$ silicon structure with an integrated DEPFET sensor (Fig.~\ref{fig:module_layout}). The different pixel pitches are listed in Tab.~\ref{tab:pitches}.
The End Of Stave (EOS) and frame of the module are \SI{525}{\micro\metre} thick.
For read-out and sensor operation, there are 14 ASICs bump-bonded to it.
Only the sensor and the balcony with the Switcher ASICs on it are located within the tracking acceptance of the detector while the EOS and read-out ASICs are outside and directly mounted onto cooling blocks \cite{Ye:2018}.
Three layers of metal lines are included in the HLL DEPFET process~\cite{Richter2003} and provide the connections between sensor pixels, ASICs and surface mounted passive components on the top side of the module.
The connection to the back-end electronics and power is implemented by a kapton flex, which is soldered and wirebonded to the EOS.

\begin{figure}[htbp]
  \centering
  \includegraphics[width=0.7\textwidth]{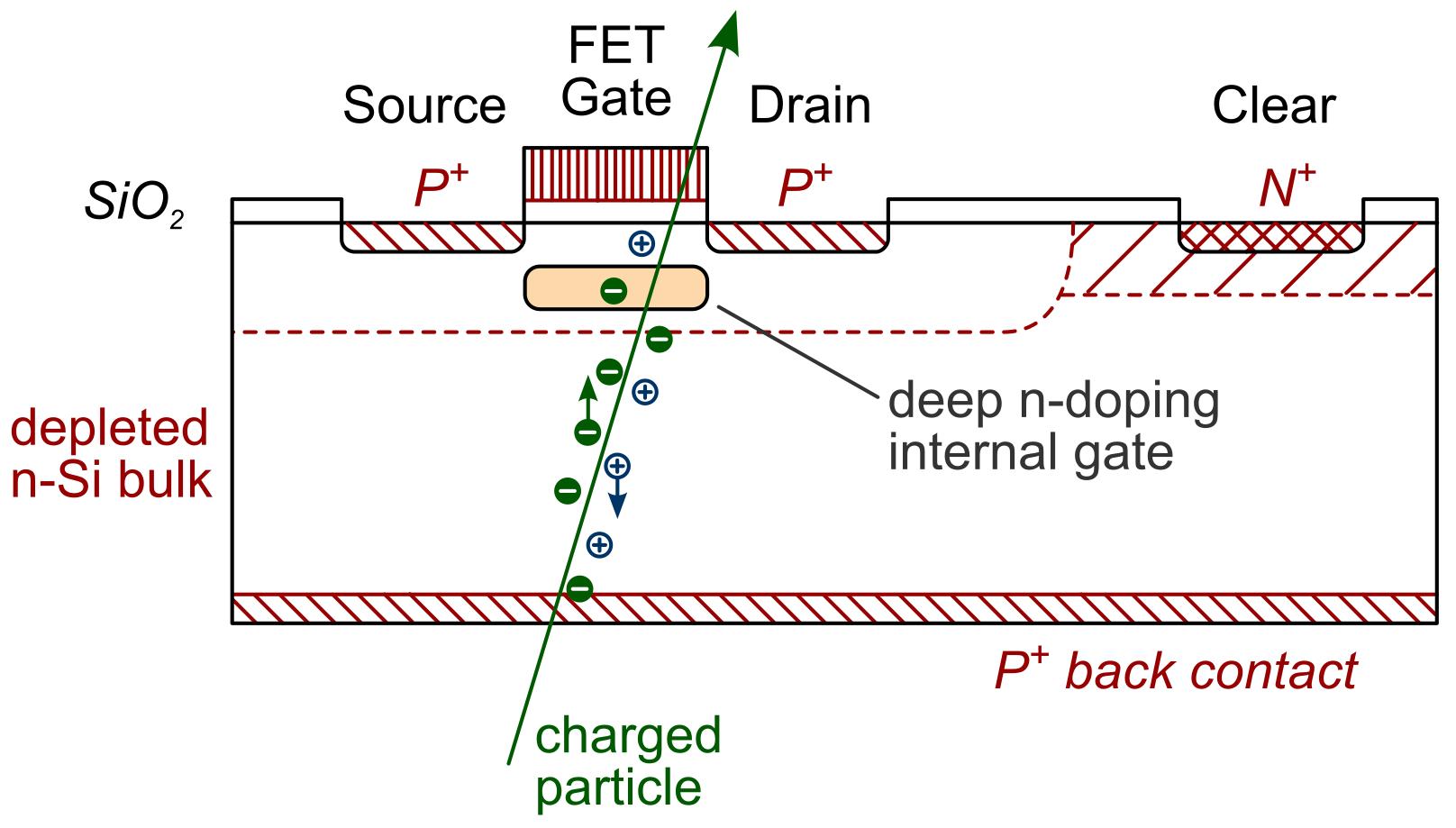}
  \caption{Schematic cross-section of a DEPFET pixel cell \cite{Richter2003}.}
  \label{fig:DEPFET_schema}
\end{figure}

% DEPFET sensor
The sensor consists of a $250\times768$ matrix of DEPFET pixels. It is thinned to \SI{75}{\micro\meter} to reduce the average material budget to $\approx \SI{0.2}{\percent} X/X_0$ per layer inside the tracking acceptance.
Each pixel features a FET on top of the depleted silicon bulk (Fig.~\ref{fig:DEPFET_schema}).
The bulk depletion grows from the front- and backsides forming a potential maximum (PM) at their intersection.
The PM is enhanced by a deep n implantation forming the so-called internal gate.
The holes generated by ionising radiation drift to the backside contact.
The free electrons are collected in the internal gate, just under the FET. Typical charge collection times are in the order of \SI{10}{\nano\second}~\cite{Schwenker:2014}.
The presence of the negative charge $ Q$ influences the channel of the FET and increases the source-drain current $I_D$ by $g_q \times Q$, which serves as the signal.
During the time of signal integration, the FET is switched off by applying a voltage well above the threshold.
When the FET is switched on by setting its gate to a negative voltage ($\sim\SI{2}{\volt}$), the drain current signal can be sampled.
After signal sampling the charges are removed by applying a positive voltage of \SI{20}{\volt} to an additional n+ implant in the pixel (labelled 'Clear' in Fig.~\ref{fig:DEPFET_schema}) while the negative gate voltage is still applied.
During the time of the clear pulse, a conducting channel is formed between the internal gate and the clear implant so that the electrons drift to the positive contact.
The gate and clear contacts of every four consecutive rows of pixels are connected together forming segments of 1000 pixels along the matrix which are switched simultaneously.
This results in 1000 drain lines each connecting the drains of 192 pixels, one per four-row segment, to the digitiser chip.
For matrix read-out, the four-row segments of the matrix are sequentially switched on for $\sim\SI{100}{\nano\second}$ each.
This results in a rolling shutter read-out with a period of \SI{20}{\micro\second}.
The read-out of each pixel thus takes \SI{0.5}{\percent} of the full integration time. The clearing period is \SI{20}{\nano\second} for each read-out group, which translates to \SI{0.1}{\percent} of the full integration time. It is not fully investigated yet to what extent the hit efficiency is affected when a particle hits a pixel while clearing. 

% ASICs
% Switcher
The pulsed operation of the FET gate and clear voltages of the different read-out groups is steered by the six Switcher ASICs~\cite{DCDBSwitcher} on the balcony alongside the module.
% DCD
The read-out of the drain currents is implemented in four Drain Current Digitizer (DCD) ASICs~\cite{DCDBSwitcher} at the end of the module.
Each DCD features 256 channels with dedicated Transimpedance Amplifiers (TIAs) with a globally adjustable gain at the input stage.
The amplified signals are digitised by 8-bit pipeline Analog Digital Converters (ADCs). Signals are measured in arbitrary digital units (ADU).
% DHP
The clock distribution and sequencing of the DCDs and Switchers is carried out by four Data Handling Processor (DHP) ASICs~\cite{KRUGER2010337}.
They produce a \SI{305}{\mega\hertz} clock for the on-module communication and receive the digitised values of the drain currents from the DCD.
When a trigger is received, each DHP sends out zero-suppressed hit data via a \SI{1.6}{\giga\hertz} link.
The triggered data read-out is selected for one four-row segment at a time by a read pointer which is increased in step with the rolling shutter.
To receive a full matrix frame containing the data of all hit pixels at an instant $t_0$, the trigger signal has to be applied for \SI{20}{\micro\second}. The full matrix frame contains hit pixels of a time span of $\pm$\SI{20}{\micro\second} around $t_0$. The read-out can be triggered for an arbitrary length of time. The length of the trigger signal is usually chosen with an additional margin to account for uncertainties.

% Explanation of pedestals
For hit detection, the digitised pixel signals are compared to their stored pedestal values in the DHP memory and an adjustable threshold is applied.
The pedestal values are computed as the averaged digitised drain current. They are measured before data taking periods by reading out the raw digitised current values of 100 read-out cycles and averaging them.
This raw data read-out is a special read-out mode offered by the DHP, which dumps the data memory contents without any zero-suppression. The noise of the pixel signal is defined as the standard deviation of the raw values of the pedestal measurement.
All drain current values have to be within the dynamic range of the ADCs. Because of process variations in the sensors and the DCDs there is a relative spread in the measured pedestals of the individual pixels, which limits the maximum applicable gain of the TIA in front of the ADC input.

\begin{figure}[p]
  \centering
  \includegraphics[width=\textwidth]{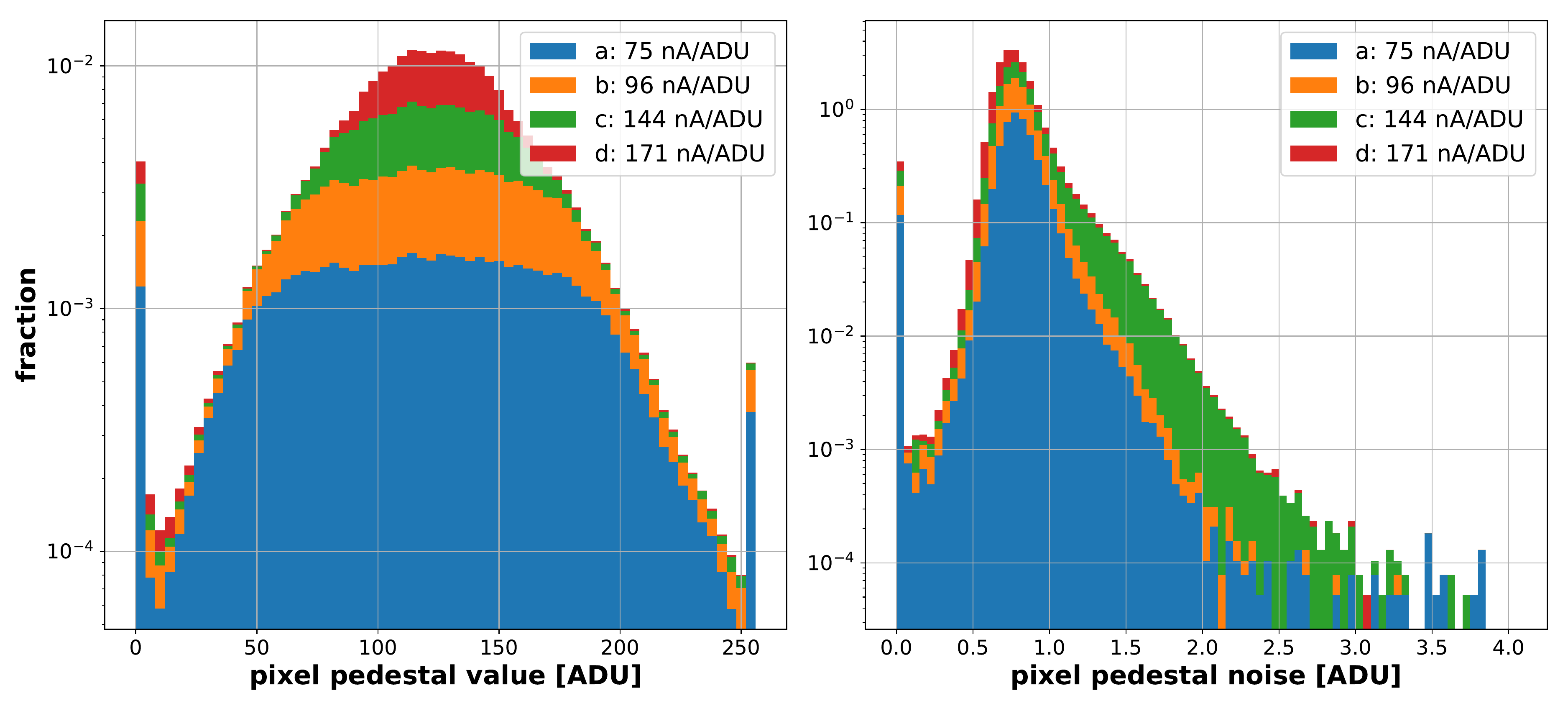}
  \caption{\textit{(colour online)} \textit{left}: Typical distribution of the digitised pedestal currents of all $250\times768$ DEPFET pixels of module \WelevenOFtwo{} as a function of the TIA gain (stacked histogram). The $y$-axis is number of pixels normalised to unity (divided by 192,000) for each distribution individually. \textit{right}: Noise distribution for all pixels of the same module. The $y$-axis is number of pixels normalised to unity for each distribution individually.}
  \label{fig:pedestal_distribution}
\end{figure}

Figure~\ref{fig:pedestal_distribution} shows the pedestal distribution and noise of a DEPFET module after optimisation for the four largest gain settings.
The pedestal distribution is centered in the dynamic range with only a small fraction of functioning pixels close to the edges.
The entries in the bottom and top bins in this case are due to a number of bad pixels from known matrix level or switcher channel defects in this module.
The right plot shows that the noise is largely independent of the amplifier gain and therefore is dominated by electronic noise in the DCDs.
The average pedestal noise is in the order of $\SI{0.7}{ADU}$.
This also means that a higher DCD gain increases the signal-to-noise ratio.
The entries at 0 stem from the bad pixels mentioned above.

\begin{table}[p]
    \centering
    \begin{tabular}{c|c|c}
         &  inner area pitch ($u\times v$) & outer area pitch ($u\times v$)\\
        \hline
        L1 & $\SI{50}{\micro\metre}\times\SI{55}{\micro\metre}$ & $\SI{50}{\micro\metre}\times\SI{60}{\micro\metre}$ \\
        L2 & $\SI{50}{\micro\metre}\times\SI{70}{\micro\metre}$ & $\SI{50}{\micro\metre}\times\SI{85}{\micro\metre}$ \\
    \end{tabular}
    \caption{Pixel pitches of the different PXD modules in the first (L1) and second (L2) layer, for the outer and inner sensor matrix area, respectively.}
    \label{tab:pitches}
\end{table}

\section{DEPFET detector setup}
% setup overview
For the presented measurements the modules were operated with the same components as used in Belle~II.
% PS
The power for the ASICs and all matrix terminals and implants is provided by a custom-built 24 channel Power Supply (PS)~\cite{RUMMEL201351}.
The PS features remote voltage sensing for every channel and is integrated into the PXD Experimental Physics and Industrial Control System (EPICS) slow control \cite{Ritzert2015}.
% DHE
Module configuration and read-out are performed by an AMC based FPGA board, the Data Handling Engine (DHE), with a laboratory specific firmware.
The DHE is integrated into the slow control via IPBus.
Events are built from the received data and sent out via UDP to a computer where they are recorded with a local DAQ program. A full description of the Belle~II PXD back-end electronics can be found in~\cite{levit2015fpga}.
% slow control
The module control is handled with the EPICS based production slow control.

\section{Beam test setup}
The beam tests were conducted at the DESY test beam facility~\cite{desy-tb:2019} which provides a mono-energetic electron beam with a tunable energy between \SI{2}{\giga\electronvolt} and \SI{6}{\giga\electronvolt}. A EUDET-type beam telescope was used for the measurements~\cite{Jansen:2016}. It consists of six Mimosa26 high-resolution Monolithic Active Pixel Sensor (MAPS) planes with \num{1152 x 576} pixels and $\SI{18.4}{\micro\metre}\times\SI{18.4}{\micro\metre}$ pitch per pixel~\cite{HUGUO:2010}. Both the Mimosa26 and the DEPFET modules are read out in a rolling shutter mode. The integration time of the Mimosa26 is \SI{115.2}{\micro\second}~\cite{Jansen:2016}, while the DEPFET module's integration time is \SI{20}{\micro\second}. Therefore, an additional detector plane providing precise timing information was employed. A \num{336 x 80} pixels ($\SI{50}{\micro\metre}\times\SI{250}{\micro\metre}$ pitch) hybrid pixel detector with an FE-I4~\cite{GarciaSciveres2011} front-end read-out chip and \SI{25}{\nano\second} time resolution was placed downstream after the last telescope plane~\cite{Obermann_2014}. This telescope plane will be referenced simply as FE-I4 in this paper.
The FE-I4 also provides a trigger signal which is issued once any pixel has a signal above threshold.
The read-out of all detectors was triggered by a coincidence of the FE-I4 plane and a scintillator in front of the telescope.
The EUDET/AIDA Trigger Logic Unit (TLU) performed the coincidence determination and trigger distribution.

The length of the Mimosa and DEPFET read-out frames and their timing with respect to the trigger is chosen such that the hits of the triggering particle are within their respective signal integration times.
Any reconstructed particle track that has associated hits on the six Mimosa26 and the FE-I4 planes must have traversed the DEPFET module within its integration time window. These tracks are used for determining the hit efficiency of the DEPFET module. 

The online data acquisition for all detectors was integrated into the EUDAQ framework~\cite{EUDET2020}. EUDAQ serves as a central run control and data collector for the individual detectors' DAQ systems. Read-out frames of each detector are collated in a single EUDAQ event with associated trigger number $N_T$. Consecutive events are collected in a run file. Data taking was structured in runs of 250k to 500k events for our beam tests.

It was observed that in about every second run, a mismatch of the trigger number among the individual detectors' hit data frames within an event was present. This was corrected by reassembling the individual hit data frames with matching trigger numbers.

\begin{figure}[p]
    \centering
    \includegraphics[width=.99\textwidth]{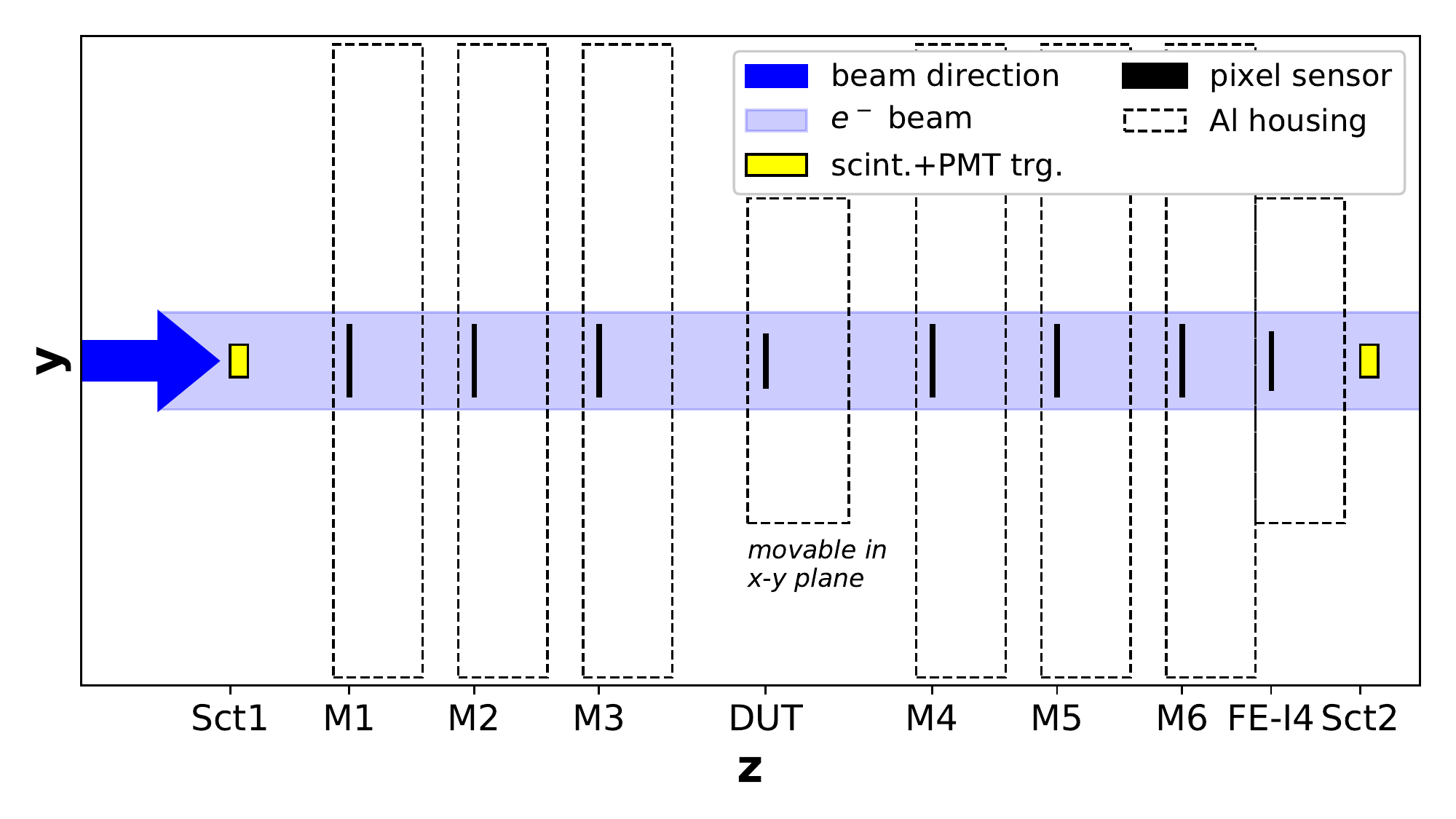}
    \caption{\textit{(colour online)} Sketch of the geometry of the beam test setup. The dimensions are given with correct relations. The $z$-axis is identical to the beam axis, the $y$-axis is vertical and the $x$-axis is going out of the sketch plane. M1 to M6 label the six Mimosa26 pixel devices of the beam telescope. The DUT is placed between the two arms of the beam telescope. Two scintillators, read out with PMTs, are placed upstream and downstream (Sct1 and Sct2) of the beam. The DUT is mounted on motor stages allowing for precise movement within the $x-y$ plane.}
    \label{fig:testbeam_geometry}
\end{figure}

\paragraph{Telescope geometry}
Figure~\ref{fig:testbeam_geometry} shows the geometry of the beam test setup. There were two telescope arms with three Mimosa26 sensor planes each, and the FE-I4 reference plane in the downstream arm. Details about the support and cooling infrastructure of the beam telescope can be found in~\cite{Jansen:2016}. The Device Under Test (DUT), the DEPFET module, was mounted on an aluminium cooling block, kept at $10\,^\circ\mathrm{C}$ by a water chiller, with cutouts below the sensor area, and placed inside an aluminium housing. Nitrogen was flowed inside the housing to prevent condensation.
The geometry and coordinate system of the beam test setup is shown in Fig.~\ref{fig:testbeam_geometry}. Values given in DUT coordinates use the coordinate system indicated in Fig.~\ref{fig:module_layout}. The DUT was mounted on motor-stages allowing for movement in $x$- and $y$-direction. Additionally, a motor-stage for precise rotation around the $x$-axis was employed. Rotations around the $y$-axis were realised by manually tilting the whole DUT mounting structure. The distance between the telescope planes $D^\mathrm{tel}$ was \SI{2.1}{\centi\metre} when measuring perpendicular particle incidence on the DUT, and \SI{5}{\centi\metre} when measuring inclined particle incidence. The distance between the two arms $D^\mathrm{DUT}$, where the DUT is placed, was \SI{5.8}{\centi\metre} for perpendicular incidence and up to \SI{30}{\centi\metre} for inclined incidence, due to the dimensions of the DUT housing. For perpendicular incidence, a beam energy of \SI{3}{\giga\electronvolt} was used, while it was increased to \SI{5}{\giga\electronvolt} when measuring inclined particle incidence with larger spacing. Values for $D^\mathrm{tel.}$, $D^\mathrm{DUT}$ and beam energy for the various incidence angles are listed in Table~\ref{tab:hit_prediction_precision}.

\paragraph{Devices under test}
In total four production PXD modules were tested during two beam test campaigns in November 2018 and April 2019. All four pixel designs are covered by the DUT selection. All DUTs were manufactured in a common production run but stem from different silicon wafers. Thus, the variation of the drain currents across all DEPFET pixels is very module specific and involves an adapted tuning of operation parameters. The read-out ASICs are identical. One module (\WfiveOBone) was measured before and after (denoted as \WfiveOBoneIrrad) irradiation with soft X-ray photons to a total ionising dose of about \SI{266}{\kilo\gray}~\cite{Schreeck2020} of the sensor oxides. Table~\ref{tab:pxd_sensor_modules} summarises the key performance numbers of the tested sensor modules as measured in laboratory characterisations~\cite{Wieduwilt:2016}.
The pedestal spread $\Delta_\mathrm{ped}$ is measured as the FWHM of the pedestal distribution, excluding the lowest and highest bin which collect unusable pixels (cf.~Fig.~\ref{fig:pedestal_distribution}). Different tuning of the DEPFET gate potential, $V_\mathrm{gate}$, and the digitisation gain on each module is necessary for a comparable pedestal spread. The reason is a module specific drain current distribution as a result of the production process. The fraction of alive pixels is a measure for the sensor module quality. It is the fraction of pixels that show a non-zero signal response, when the sensor is exposed to a radioactive Strontium-90 source, over all $250 \times 768$ pixels. The DUTs \WelevenOFtwo\ and \WfiveOBone\ are grade-B with $<99\,\%$ alive pixel. Moreover, the \WfiveOBone\ has one out of four read-out ASIC pairs not functioning. After irradiation, the gate potential has to be significantly more negative for \WfiveOBoneIrrad\ due to the oxide damage. Also, a few read-out rows were damaged during the irradiation campaign on \WfiveOBoneIrrad\ due to an operator mistake. A gain setting of \SI{96}{\nano\ampere\per ADU} is used, as used for the PXD operated in Belle~II, when possible. In the beam tests it is used for all DUTs except \WfourtyIF, which is operated with lower amplification at a gain setting of \SI{144}{\nano\ampere\per ADU} due to a larger intrinsic spread of the pedestal distribution.
The zero-suppression threshold used in the beam tests is $>\SI{4}{ADU}$, except for the irradiated module \WfiveOBoneIrrad\ which was operated with a threshold of $>\SI{5}{ADU}$.

\begin{table}[p]
\centering
\begin{tabular}{l|c|c|c|c|c|}
 DUT & $\Delta_\mathrm{ped}$ [ADU] & noise [ADU] & alive [\%] & $V_\mathrm{gate}$ [V]& gain [nA/ADU]\\
 \hline
 \WfourtyIF    & 104 & $0.8\pm0.2$ & $99.1$ & -1.6 & $144\pm15$ \\
 \WelevenOFtwo & 115 & $0.8\pm0.2$ & $97.4$ & -2.0 & $96\pm10$ \\
 \WfiveOBone   & 92 & $0.6\pm0.2$ & $74.2$ & -2.1 & $96\pm10$ \\
 \WfiveOBoneIrrad & 162 & $0.6\pm0.2$ & $73.1$ & -10.7 & $96\pm10$ \\
 \WphaseTwo       & 84 & $0.7\pm0.2$ & $99.8$ & -4.0 & $96\pm10$\\
\end{tabular}
\caption{Properties of the DUTs investigated in the beam tests as measured in previous laboratory characterisations.}
\label{tab:pxd_sensor_modules}
\end{table}

\FloatBarrier

\section{Event reconstruction}
\label{sec:track_reco}
\begin{figure}[p]
  \centering
  \includegraphics[width=.75\textwidth]{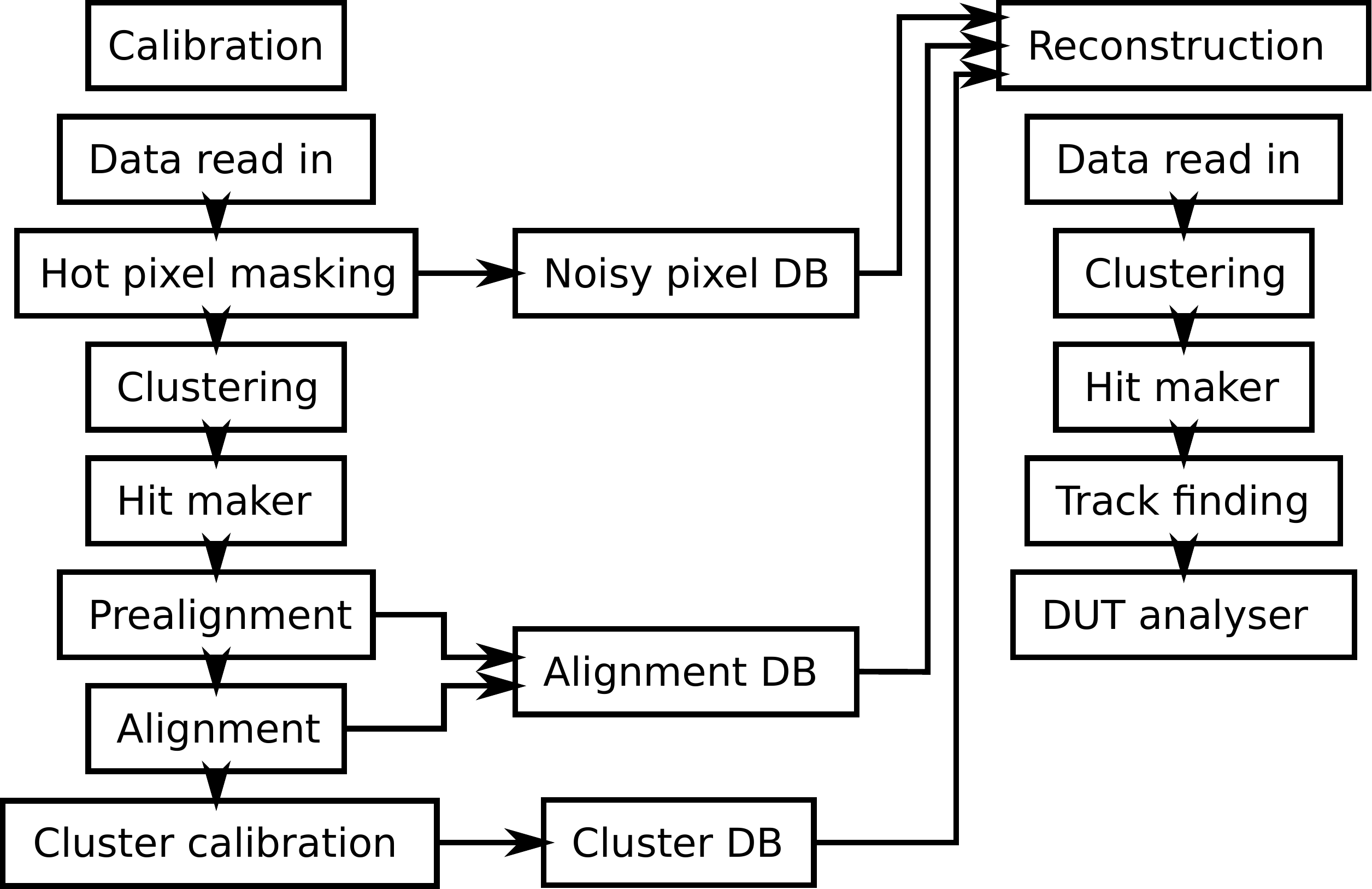}
  \caption{Data calibration and track reconstruction flow chart sketching the individual processing steps. The calibration is performed on a subset of the data producing pixel masks and alignment parameters, which are used in the reconstruction to produce tracks for the full dataset.}
  \label{fig:track_reco_flow}
\end{figure}

The reconstruction of electron tracks through the telescope and the extrapolation of hit positions onto the DUT is done with the Test Beam Software Framework (TBSW)\cite{tbsw}. The general process flow is shown in Fig.~\ref{fig:track_reco_flow} and described in more detail in~\cite{Schwenker:2014, Stolzenberg:2020}. The reconstruction of an event begins with clustering the digitised and zero-suppressed measurements from all sensors. The clusteriser removes signals from defective pixels using a pre-computed hot pixel mask and groups pixels sharing a common edge or corner on the sensor into a pixel cluster. The position finding step computes an estimate of the 2D hit position and its $2\times2$ covariance matrix for each cluster on each sensor. Afterwards, hits on multiple telescope sensors are combined to tracks in the track finding step and a Kalman filter is run to estimate the particle's intersection coordinates and incidence angle on every sensor. The final DUT analysis requires as input a sample of tracks found in the reference telescope (without using hits from the DUT) as well as clusters on the DUT. Tracks are required to have hits on at least five of the six Mimosa26 telescope sensors. A hit on the FE-I4 plane is not required but will be added to the track if present. For each track, the predicted intersection on the DUT sensor is computed and the closest DUT cluster in the same event is found. This DUT cluster is matched to the track if the distance is smaller than \SI{200}{\micro\metre} along the $u$- and $v$-coordinate. The output is a pair of tables for clusters and track intersections on the DUT sensor together with their matching relations. Calibration constants are required at different steps of the reconstruction and are computed from dedicated calibration runs before the reconstruction of data can be started. Three different kinds of calibrations are required: Hot and dead pixel masks (noiseDB), alignment corrections to the position of sensors (alignmentDB) and a lookup table for mapping pixels clusters to 2D hit positions (clusterDB). The next sections provide more information regarding the different algorithms.   

\begin{figure}
    \centering
    \includegraphics[width=0.85\textwidth]{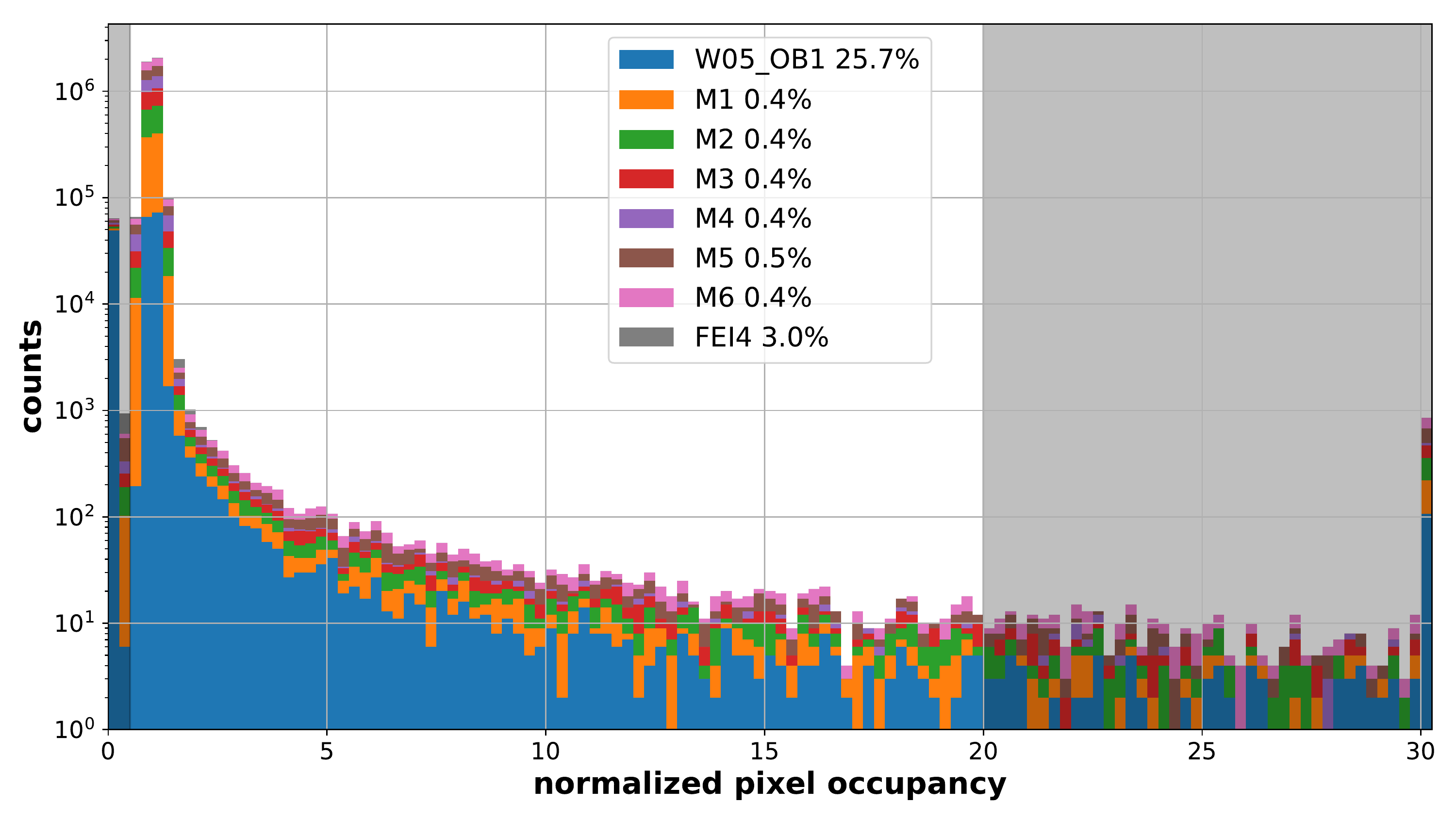}
    \caption{\textit{(colour online)} Distribution of normalized pixel occupancies $\Omega$ for the DUT, the FE-I4 timing plane and the six Mimosa26 (M1 to M6) telescope planes. Pixels with occupancies within the grey areas are masked as dead and hot pixels. The legend lists the ratio of masked pixels for each sensor plane. The hot pixel threshold was chosen such that only a permille of pixels are rejected from further analysis. The zero occupancy bin for the DUT \WfiveOBone\ is large, since one of four ASIC pairs ($\widehat{=}\,\SI{25}{\percent}$) is not functioning (cf.~Tab.~\ref{tab:pxd_sensor_modules}). The highest occupancy bin is an overflow bin. The distributions are stable over time for the respective sensors.}
    \label{fig:occupancy_distribution}
\end{figure}

Hot and dead pixel masks are computed for all sensors in the telescope using calibration runs with 250k to 500k events. For each pixel on a sensor, the masking algorithm first counts the number of hits over the entire calibration run. The normalized occupancy $\Omega$ of a pixel is computed as the ratio of the hit count in that pixel divided by the median hit count in a $5\times5$ neighbourhood centred on that pixel. Dead pixels can be identified as having $\Omega<t_\text{dead}$, hot pixels as having $\Omega>t_\text{hot}$. The thresholds $t_\text{dead} = 0.5$ for dead pixels and $t_\text{hot} = 20$ for hot pixels were used. Figure~\ref{fig:occupancy_distribution} shows the distribution of normalized pixel occupancies per sensor plane. Indicated are the exclusion regions. The advantage of the normalized occupancy approach is that the masking can be applied to runs recorded with enabled beam. With the given thresholds, pixels with either very low or an abnormal high hit count are selected. The fraction of masked pixels is in the order of a few permille.

\paragraph{Kalman filter for track fitting}
Kalman filters are employed for track finding and fitting \cite{Fruehwirth1987}. For a linear dynamic system with Gaussian measurement and process noise, the Kalman filter is the optimal recursive estimator of the state vector~\cite{Kalman1960}. In the case of track fitting, the dynamic system is the evolution of the state of a particle traversing the telescope planes. The state at plane $k$ is described by a vector 

\begin{equation}
\lambda_{k} = (\tan \phi, \tan \theta , u, v) \label{eq:forward_state}
\end{equation}   
where $\tan \phi = du/dw$ and $\tan \theta = dv/dw$ are the local direction tangents and $u$ and $v$ are local intersection coordinates on plane $k$. Since there is no magnetic field, the default particle hypothesis is that the tracks are electrons with momenta equal to the beam energy. The track model $\lambda_{k}=f_{k|k-1}(\lambda_{k-1}, \varphi_{k-1}, \vartheta_{k-1})$ propagates the track state from sensor to sensor along a 3D straight line. The Coulombic scattering of the particle at sensor $k-1$ is described by a pair of projected scattering angles $\varphi_{k-1}$ and $\vartheta_{k-1}$ in the co-moving frame of the particle with covariance matrix $Q_{k-1}$ computed as described in~\cite{Wolin1983}. 

The first part of the Kalman filter propagates an estimate of the track state at sensor $k-1$ to $k$ and yields a predicted track state $\lambda_{k}^\mathrm{pred}$ and its covariance matrix $C_k^\mathrm{pred}$ on plane $k$

\begin{equation}
\lambda_{k}^\mathrm{pred} = F_{k|k-1} \lambda_{k-1} \ , \label{eq:forward_state_vec}
\end{equation}   

\begin{equation}
C_{k}^\mathrm{pred} = F_{k|k-1} ( C_{k-1} + G_{k-1}Q_{k-1}G_{k-1}^T   )F_{k|k-1}^{T} \ .   \label{eq:forward_state_cov}
\end{equation}   

Here $F_{k|k-1}$ is the Jacobian matrix of the track model $f_{k|k-1}$ with respect to the track parameters. The second term takes into account the effect of multiple scattering at plane $k-1$ and the formulas for computing $G_{k-1}$ are found in~\cite{Wolin1983}. The second part of the Kalman filter updates the predicted state with a noisy measurement of the track intersection coordinates provided by the sensor $k$. The 2D position measurement is $m_k$ and the measurement noise is approximated by a 2D normal distribution with $2\times2$ covariance matrix $V_k$. The details for computing hit positions and their covariance matrix will be covered in the section on position finding algorithms below. If there are multiple hits on plane $k$, the hit closest to the predicted intersection on the plane is selected for the update. The quantity describing the distance between the hit position and the predicted intersection is the residual $r_k$

\begin{equation}
r_{k} = m_k - H \lambda_k \ ,    \label{eq:residual_vec}
\end{equation}   

\begin{equation}
R_{k} = V_k + H C_k^\mathrm{pred} H^T \ .    \label{eq:residual_cov}
\end{equation} 

The constant $2\times4$ matrix $H$ projects the local track state onto the local hit coordinates. The residual covariance matrix, $R_k$, is the sum of the hit covariance matrix $V_k$ and the predicted pointing resolution of the track  $HC_{k}^\mathrm{pred}H^T$ using downstream hits only. The measurement error of the hit at plane $k$ and the prediction error of the intersection coordinates of the track using hits from upstream sensors $k
^{\prime}<k$ are taken to be fully uncorrelated. Hit candidates for updating the track state are selected by requiring the residuals to be less than \SI{400}{\micro\metre} along the $u$- and $v$-axes. The variable  $\chi^2_{inc,k}$ computed as $r^{T}_{k}R_{k}^{-1}r_{k}$ measures the statistical compatibility of a candidate hit with the predicted track state. After alignment of the telescope (see section below), a cut $\chi^2_{inc}<20$ is employed to filter outlier hits. Once a hit candidate is selected, the Kalman gain matrix method~\cite{Fruehwirth1987} is used to compute an updated track state $\lambda_k$ and track covariance matrix $C_k$. 

The Kalman filter prediction and update steps are used recursively until the hit on the last telescope sensor is reached. The final track $\chi^2$ is computed as the sum of $\chi^2_{inc,k}$ for all hits added to the track. After alignment, only tracks with a  $\chi^2<100$ are kept. A backward pass of the Kalman filter is applied yielding predictions of the track state at all planes $k$ using only hits from downstream sensors ($k^{\prime}>k$). The final prediction for the track state at a sensor is computed as the weighted mean of the predictions from both passes of the Kalman filter. In this way, the errors of the predicted intersection position from the track fit at sensor $k$ did not use a measurement at sensor $k$ and the covariance matrix of the hit track residuals can be computed as in Eq.~\ref{eq:residual_cov}. 

For the telescope alignment, the silicon sensors are treated as rigid bodies and their position and orientation are described by the coordinate transformation of a space point $r=(x,y,z)$ from global telescope coordinates to local sensor coordinates $q=(u,v,w)$. The transformation law $q=R(r-r_0)$ contains the origin of the sensor $r_0$ and a $3\times3$ rotation matrix $R$. Following the approach presented in~\cite{Karimaki}, alignment corrections are described by a shift vector, $\Delta r$, and a small rotation matrix, $\Delta R$, and are applied as $r_0 \to r_0 + \Delta r$ and $R \to R\Delta R$. In the approximation of small tilts, the rotation $\Delta R$ is constructed as the product of three individual rotations with angles $\Delta \alpha$, $\Delta \beta$ and $\Delta \gamma$ and there are six alignment parameters per sensor in total. Firstly, the position of the first sensor is fixed and shift corrections $(\Delta x, \Delta y)$ for all downstream sensors are computed by converting hits into global $x,y,z$ coordinates and looking at the shifts of residual histograms along the  $x$ and $y$ axis between hits on the reference sensor and hits on downstream sensors. After updating the sensor positions with these shift corrections, the alignment is good enough to find a sample of tracks leaving hits on all sensors in the telescope including the DUT. The final alignment parameters are computed by minimising a global $\chi^2$ function,

\begin{equation}
\chi^2 = \sum_i^\mathrm{tracks} \sum_j^\mathrm{hits} r_{ij}^T R_{ij}^{-1}r_{ij} + (a-a_0)^T W_0 (a-a_0) \ ,
\label{eq:global_chi2}
\end{equation} 
where the indices $i$ and $j$ run over all tracks and hits in the sample, $r_{ij}$ are the residuals and $R_{ij}$ residual covariance matrices. The vector $a$ concatenates six alignment parameters from each sensor and $a_0$ and $W_0$ are the initial alignment parameters and the inverse of their covariance matrix. Specifically, the matrix $W_0$ is taken to be diagonal with zero entries for the parameters to be kept fixed during alignment. The alignment parameters of the first and last Mimosa26 sensors are fixed to zero. Also, the rotations of sensors around the $x$ and $y$ axis are fixed with the exception of the DUT sensor where all three rotations angles are fitted. The Kalman Alignment Algorithm with Annealing~\cite{Fruehwirth2003} is a sequential method derived from the Kalman Filter to minimise the global $\chi^2$ function with respect to track parameters and alignment parameters. It is a sequential method in the sense that it processes tracks one by one and updates the alignment parameters and their covariance matrix after each processed track. The residuals of the next track are re-computed in the telescope geometry with updated alignment. The minimisation of $\chi^2$ in Eq.~\ref{eq:global_chi2} is iterated with increasing quality cuts on the track sample. In the first iteration, hits are added to tracks with loose cuts on the residuals ($<\SI{500}{\micro \metre}$) and no cuts on the $\chi^2$ of the track is applied. In the next iterations, the cut on the residuals is tightened ($<\SI{400}{\micro \metre}$) and cuts on $\chi^2_{inc}$ ($<20$) and total $\chi^2$ ($<100$) are applied. The telescope alignment is repeated whenever the arms of the telescope are moved or the DUT sensor is rotated. 

\paragraph{Position finding algorithms}
For the position finding, the pixel signals belonging to a single particle have to be identified. This step is called clustering and the algorithm used is explained in~\cite{Moll2019PhdThesis}. All pixel sensors in the telescope feature a zero-suppressed read-out producing a list of fired pixels or digits per sensor. Starting with the pixel in the upper left corner of the sensor, the clusteriser checks each pixel for being masked as hot during the calibration. If it is not masked, the left neighbour in the same pixel row and the direct neighbours in the previous row are investigated. If one or more clusters have already been found in those neighbouring pixels, the clusters are merged and the pixel is assigned to this cluster. 
Otherwise, a new cluster is created and the pixel becomes its first member. The  clusteriser proceeds with the pixel to the right of the current pixel or, if the pixel is the last pixel in the current row, with the first pixel of the next row. The procedure is repeated until the last pixel in the last row has been processed. 

A position finding algorithm (PFA) takes the pixel cluster $c$  on a sensor as input and outputs a pair of measured hit coordinates $m=(u^{m},v^{m})$ along with a symmetrical $2\times2$ matrix $V$. The matrix $V$ is an estimate of the unknown covariance matrix of the error between measured hit position and the true particle intersection. From an experimental point of view, the best approximation for the unknown measurement error is the  predicted residual, as defined in Eq.~\ref{eq:residual_vec}, for a sample of tracks. A useful position finding algorithm yields an unbiased distribution of residuals with a standard deviation consistent with the residual covariance according to Eq.~\ref{eq:residual_cov}. 

\paragraph{Position finding by centre of gravity}
The centre of gravity algorithm is a well studied algorithm for silicon strip detectors~\cite{Turchetta1993} but can be applied to pixel detectors as well. In the context of pixel detectors, the $n$ pixels of a cluster can be indexed such that $u_i$ ($v_i$) is the $u$ ($v$) position and $q_i$ the signal of the $i^\mathrm{th}$ pixel in the cluster. The hit position $u^m$ is given by

\begin{equation}
u^{m} = \frac{\sum_{1}^{n} q_i u_i} {\sum_{1}^{n}q_i} \ .
\label{eq:cog_pos}
\end{equation}   
and  $v^m$ is determined in an analogous way. For the centre of gravity approach, the off-diagonal entries of $V$ are taken to be equal to zero. The square root of the diagonal element $V_{uu}$ is computed according to $a \times P_u/\sqrt{12}$ where $a$ is a sensor type specific scale factor and $P_u$ is the $u$-pitch of the pixel with the highest signal in the cluster. The scale factors used are $0.8$ (DEPFET), $0.7$ (Mimosa26) and $1.0$ (FE-I4). The numbers are found by a grid search inspecting the uniformity of the $p$-value distribution over a sample of tracks. The square root of $V_{vv}$ is computed using the same scale factors but for the pixel pitch in the $v$ direction. 

\paragraph{Position finding by cluster shapes}
The cluster shape method is a new generic approach to position finding designed to jointly estimate a 2D hit position and its full $2\times2$ covariance matrix. The main idea is to compute a lookup table of position corrections for a finite number of translation invariant cluster classes called shapes. Owing to the discrete symmetries of the pixel matrix, the hit position $m(c)$ of a cluster can be written as a sum of shape correction $m(s)$ and a cluster offset $O(c)=(O_u, O_v)$. The $u$ ($v$) coordinate of the cluster offset is the minimum of the pixel coordinates $u_i$ ($v_i$) over all pixels $i$ in the cluster. The shape $s$ of a cluster $c$ is the sequence of relative pixel positions $(u_i - O_u, v_i - O_v)$ sorted in ascending $v$ coordinate. The shape $s$ is used as a key into a lookup table of key-value pairs to retrieve shape corrections $m(s)$ and $V(s)$. Loosely speaking, the shape correction $m(s)$ is the \textit{centre} and $V(s)$ measures the \textit{area} on the sensor plane that a particle must hit for creating the observed shape.    

The DEPFET pixel sensors for Belle~II have small pixels in the inner area and large pixel in the outer area, see Table~\ref{tab:pitches}. Moreover, the pixel pitches differ between sensors in the first and second layer of the Belle~II  vertex detector yielding in total four different pixel sizes. The different pixel sizes can be enumerated by a pixel type $l_i$ and the cluster shape is extended to keep track of the type of each pixel in the cluster. The simplest way to use the shape $s$ as a key into a lookup table is to convert the tuples $(u_i - O_u, v_i - O_v, l_i)$ into strings and to concatenate these strings over all pixels in the cluster. 

The DEPFET sensors also provide an 8-bit signal measurement $q_i$ for each pixel in the cluster. The pixel signals are incorporated into the shape correction via the ratio $\rho$ of the charge $q_h$ in the head and the charge $q_t$ in the tail pixel of the cluster, defined by

\begin{equation}
\rho = \frac{q_h}{q_h+q_t} \ .
\label{eq:rho}
\end{equation}

It is used as an additional input feature for computing shape corrections. For clusters with more than two pixels, a heuristic is used to select the head and tail pixel signals depending on the sign of the predicted intersection angles $\theta, \phi$ of the particle beam on the sensor plane. If the angles have the same sign, the tail (head) pixel is the lower left (upper right) pixel in the cluster. In case the signs are opposite, the tail (head) pixel is the lower right (upper left) pixel of the cluster. 

For the calibration of cluster shapes in a beam test, an initial sample of fitted tracks in an aligned telescope is required. In order to obtain such an initial track sample and telescope alignment, the centre of gravity method is used. The track selection criteria are identical to those for the last iteration of the track based telescope alignment.  For the cluster shape calibration of a DUT sensor, a dataset $D=\{c_j, \lambda_j^\mathrm{pred}, C_j^\mathrm{pred}\}$ is formed containing the state predictions for track $j$ at the DUT sensor and a matched DUT cluster. Since the particle beam provided by the DESY test beam facility is highly collimated~\cite{desy-tb:2019}, the incidence angles of tracks recorded in $D$ are defined by the rotation angles of the sensor with a very small angular spread ($<\SI{2}{\milli \radian}$). Similar to the alignment, the cluster shape calibration is repeated after rotating the DUT sensor. The mean incidence angles of particles in the calibration data are denoted $\theta$ and $\phi$. 

The dataset $D$ is scanned to produce a list of unique cluster shapes $s$ together with counts of their occurrence. If the occurrence count is below a threshold $N_{c}$, typically a few hundred, no calibration is performed and the centre of gravity method is used as the fallback position finding method for clusters of this shape. For the remaining shapes $s$, the subsets $D(s)$ with all clusters belonging to a particular shape $s$ are formed. For each subset $D(s)$ the distribution of $\rho$ values is formed and equal frequency binning is used to find a binning in $\rho$ specific for shape $s$. The default number of bins $N_{\rho}$ is a hyperparameter of the cluster shape method. The actual number of bins for a shape will be reduced to always satisfy the minimum occurrence count larger than $N_{c}$ in all $\rho$ bins. The number of bins is set to one for single pixel shapes or for sensors providing only binary signals.  

At this point, the dataset $D(s)$ can be split into subsets $D(s,\rho_i)$ of track predictions linked to a cluster with specific shape and $\rho$ bin. In order to compute shape corrections, the cluster offsets need to be subtracted from the predicted track intersections and the sample mean over all tracks in $D(s,\rho_i)$ is computed:

\begin{equation}
m( s, \rho_i )  =  \langle H \lambda_{j}^\mathrm{pred} - O(c_j) \rangle_{j \in D(s, \rho_i)} \ . 
\label{eq:lm_mean_est}
\end{equation}

This choice for $m( s, \rho_i )$ avoids a bias in the hit position. The predicted residuals can be formed and the sample covariance matrix over all tracks in $D(s,\rho_i)$ can be computed. The final step is to employ Eq.~\ref{eq:residual_cov} to estimate the covariance matrix of the shape correction:  

\begin{equation}
V( s, \rho_i )  = \textrm{Cov} (  H \lambda_{j}^\mathrm{pred} - O(c_j) - m( s, \rho_i) )_{j \in D(s, \rho_i)} - T \ ,
\label{eq:lm_pos_est}
\end{equation}

\begin{equation}
T  = \langle H C_j^\mathrm{pred} H^T \rangle_{j \in D} \ .
\label{eq:lm_cov_est}
\end{equation}

In practice, the subtraction of the telescope covariance matrix $T$ can be unstable if the diagonals of $V$ and $T$ are of similar magnitude. In particular, a systematic error on $T$ can lead to biased results for $V$. The problem can be partly mitigated by using high resolution reference sensors, a high beam energy and an optimised telescope geometry for minimising the diagonals of $T$ at the DUT sensor. 

\begin{figure}
    \centering
    \includegraphics[width=0.7\textwidth]{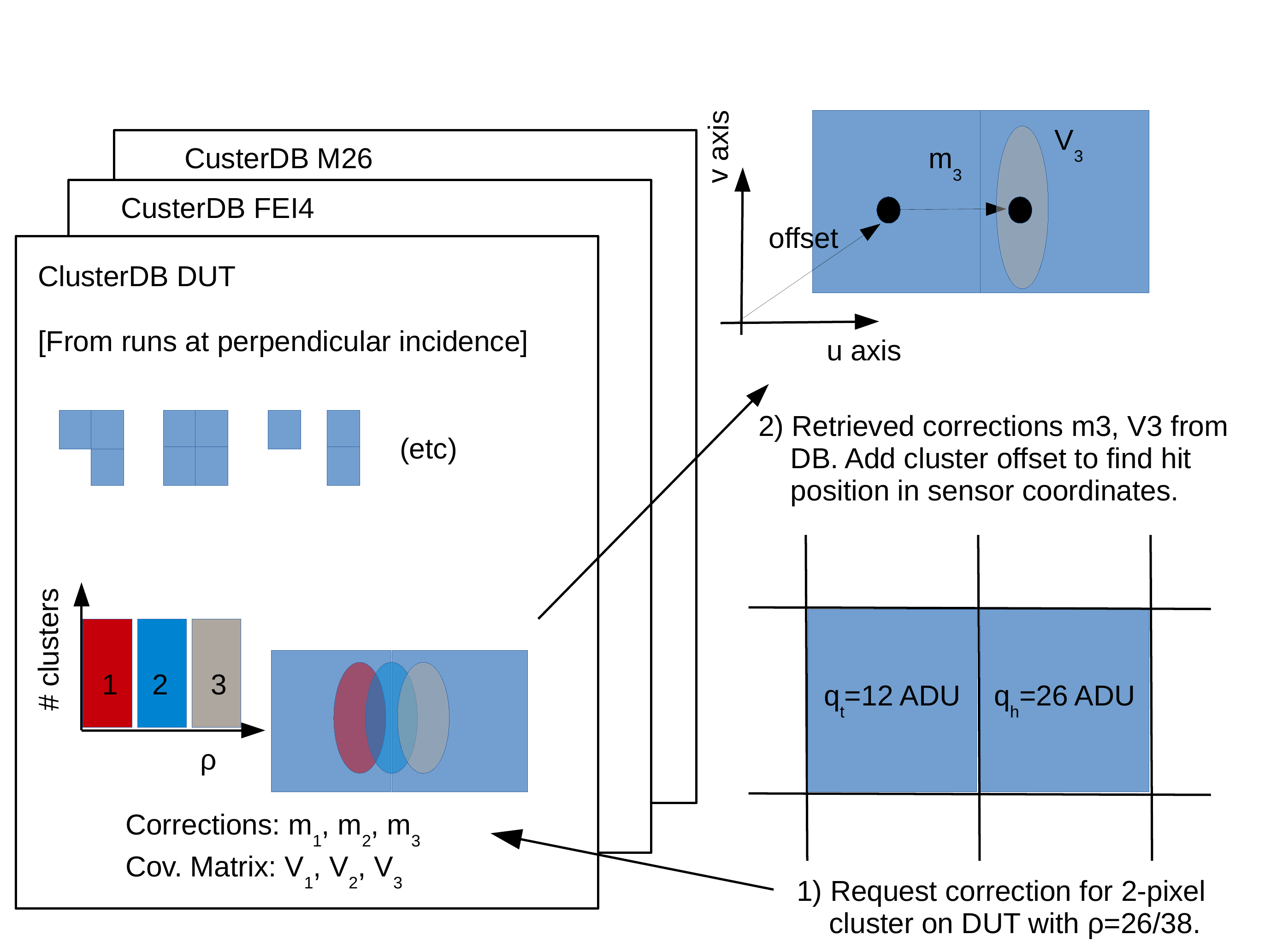}
    \caption{Illustration of the cluster shape position finding. An input cluster on the DUT sensor is processed by identifying its shape (2-pixel) and computing the ratio $\rho$ of the signals in head and tail pixels. A position corrections is requested from the DUT clusterDB. If the cluster is found in the clusterDB, the $\rho$ bin is identified (here bin 3) and the corresponding corrections $m_3, V_3$ are provided. The final hit position in sensor $u$- and $v$-coordinates is found by adding the cluster offset $O$.}   
    \label{fig:cluster_shape_examples}
\end{figure}

The cluster shape calibration can be applied to the complete telescope by repeating the procedure for all sensor types in the telescope (Mimosa26, FE-I4, DUTs). The same track sample can be used for this purpose. For the six Mimosa26 sensors, a single clusterDB is shared by all sensors but is produced only from the hit track pairs on the inner telescope planes where the telescope resolution is sufficiently high. The cluster shape calibration is iterated six times using the cluster shape corrections from the previous iteration to repeat the track fit. Figure~\ref{fig:cluster_shape_examples} illustrates the content of the clusterDB and how it is used during position finding. 

The proposed application of the cluster shape PFA for Belle~II deviates slightly from the implementation for test beams discussed above. The proposal is to use Belle~II simulations~\cite{Kuhr2019, Moll2019PhdThesis} to form the dataset $D$ from pairs of simulated PXD clusters and their related Monte Carlo truth particle intersections $(u^x, v^x)$ and incidence angles $(\tan \phi^x, \tan \theta^x)$. Next, $D$ is split into $18\times18$ bins of the incidence angles and corrections are computed separately for these bins. The cluster shape $s$ and the two possible $\rho$ values for same sign and opposite sign angles are computed during the clustering step. The retrieval of the shape corrections from the clusterDB is performed with a callback inside the track fit using the predicted track incidence angles. A full evaluation of this proposal will be given in a later publication. 

\FloatBarrier

\section{Resolution studies}
\label{sec:resolution}
The track impact parameter resolution of Belle~II is dominated by the spatial resolution of the innermost layer of the vertex detector. The DEPFET sensors in the Belle~II PXD will mostly see charged particles impinging under inclined angles. The effect of inclined particle incidence was investigated during the beam tests by independently rotating the DUT (only module \WfourtyIF\ was measured this way) around the $x$- and the $y$-axes. Data was recorded at several discrete angles. This 2D angular scan of a DEPFET module from the innermost layer is crucial for measuring the spatial resolution for inclined tracks. The cluster shape position finding algorithm (see section~\ref{sec:track_reco}) is applied to extract cluster shape specific hit position corrections and their full $2\times2$ covariance matrices separately for all measured angles. The spatial resolutions are compared with the centre of gravity position finding algorithm. 

\paragraph{Hit residuals}
The spatial resolution can be measured by comparing the cluster based hit position $(u^{m},v^{m})$ with the predicted hit position $H\lambda_{DUT}^\mathrm{pred}=(u^{x},v^{x})$ from the telescope track. The telescope track is computed without the hit from the DUT and is extrapolated to the aligned position of the DUT sensor. The covariance matrix of the hit residuals can be decomposed as 

\begin{equation}
   \textrm{Cov}( u^{m} - u^{x}, v^{m} - v^{x}  ) = V^\mathrm{int} + T \ ,
    \label{eq:residual_widths_2d}
\end{equation}
where $T$ is the $2\times2$ covariance matrix of the extrapolated hits and $V^\mathrm{int}$ is the intrinsic $2\times2$ covariance matrix of the DUT clusters. The precision of the hit extrapolation, given by the covariance matrix $T$, is $\SI[parse-numbers=false]{3.7\pm0.2 (stat.)\pm0.3(syst.)}{\micro\metre}$ in $u$- and $v$-direction for perpendicular incidence measurements and up to $\SI[parse-numbers=false]{9.0\pm0.1 (stat.)\pm0.9(syst.)}{\micro\metre}$ ($u$-direction) and $\SI[parse-numbers=false]{17.6\pm0.1 (stat.)\pm2.2(syst.)}{\micro\metre}$ ($v$-direction) for inclined incidence measurements. Table~\ref{tab:hit_prediction_precision} lists the reconstructed telescope pointing resolution for the various beam energies and incidence angles. Both beam energy and telescope geometry influence the precision of the hit prediction.

\begin{figure}[p]
    \centering
    \includegraphics[width=.49\textwidth,page=7]{residuals.pdf}
    \includegraphics[width=.49\textwidth]{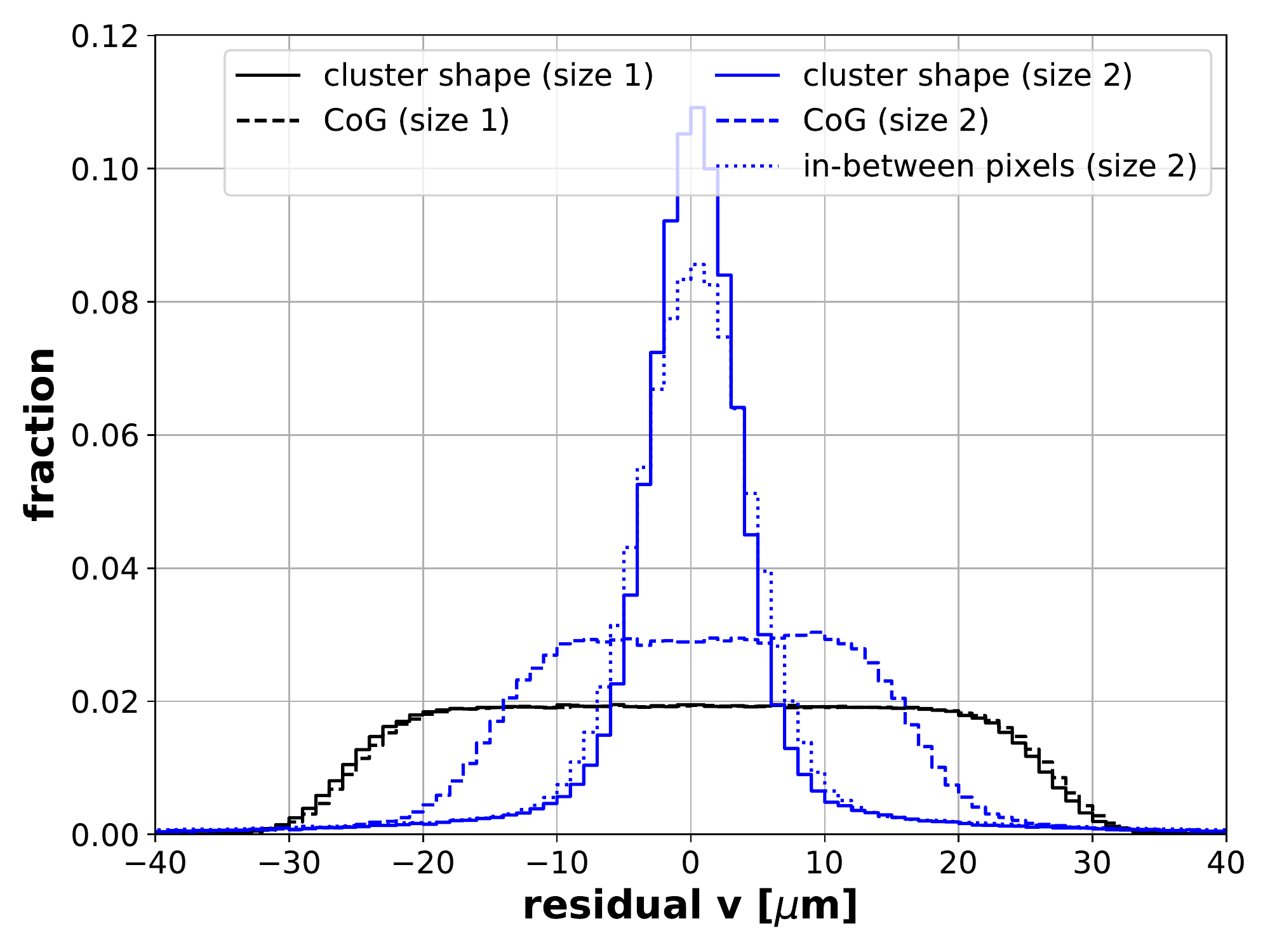}
    \caption{\textit{(colour online)} \textit{left:} Residual distribution in $u$-direction as a function of the total cluster size for perpendicular particle incidence using the cluster shape based PFA. The $y$-axis is bin counts normalised to unity for each cluster size individually.
    \textit{right:} Comparison of the residual distribution in $v$-direction for pixel pitch $\SI{50}{\micro\metre}\times\SI{55}{\micro\metre}$ at perpendicular incidence for total cluster sizes in $v$ of 1 and 2 pixels, respectively. The cluster shape PFA and CoG PFA are compared. For 2-pixel clusters, the 'in-between' line shows residuals where the hit is reconstructed in-between the two pixels. The $y$-axis is bin counts normalised to unity for each PFA individually.}
    \label{fig:residual_distribution}
\end{figure}

The intrinsic resolutions in the $u$- and $v$-direction are computed as the square root of the diagonals of $V^\mathrm{int}$, separately for the different angles. The distribution of the residuals is ideally centred at zero. A continuous drift of the residual bias over time is observed for some measurements. The bias shift is corrected by subtracting the mean residual over all tracks in each run (a set of 250k events). Figure~\ref{fig:residual_distribution} shows a corrected residual distribution for perpendicular incidence. The pixel clusters created by traversing particles are mostly one pixel clusters (\SI{83}{\percent}). There is a fraction of \SI{17}{\percent} of multi-pixel clusters with a narrower residual distribution due to charge sharing. An improvement over simple binary resolution for 1-pixel clusters is expected, since the exclusion of multi-pixel clusters reduces the effective pixel pitch. The right-hand plot in Figure~\ref{fig:residual_distribution} shows the residuals for 1-pixel and 2-pixel clusters. As expected, the 1-pixel residuals are compatible among the cluster shape based and centre of gravity PFA. However, the 2-pixel residuals for the centre of gravity method are significantly wider compared to the cluster shape method. The most likely reason is that for perpendicular incidence, the effective charge sharing region is very small compared to the pixel pitch. The centre of gravity method happens to frequently reconstruct hits at positions outside the effective charge sharing region. A cross check shows that even the simple method of always reconstructing the position exactly between two pixels gives narrower residuals in this case. To the contrary, the cluster shape corrections have 'learned' from the calibration data such that hits are reconstructed inside the charge sharing area. 

\begin{table}[p]
    \centering
    \begin{tabular}{c|c|c|c|c|c|c}
$\phi$ &  $\theta$ &  $D^\mathrm{tel}$ [\si{\centi\metre}] &  $D^\mathrm{DUT}$  [\si{\centi\metre}]&  energy [\si{\giga\electronvolt}] &  $\sigma_u^\mathrm{tel}\pm\mathrm{stat.}\pm\mathrm{syst.}$ &  $\sigma_v^\mathrm{tel}\pm\mathrm{stat.}\pm\mathrm{syst.}$ \\
\hline
 0.0 &  0.0 & 2.1 &  5.8 & 3 & $3.7\pm0.2\pm0.3$ & $ 3.7\pm0.2\pm0.3$ \\
\hline
 0.0 &  0.0 & 2.1 &  7.6 & 5 & $3.3\pm0.2\pm0.2$ & $ 3.3\pm0.2\pm0.2$ \\
 0.1 & 18.8 & 5.0 &  7.8 & 5 & $3.5\pm0.2\pm0.3$ & $ 3.7\pm0.2\pm0.3$ \\
 0.4 & 29.0 & 5.0 & 11.4 & 5 & $4.1\pm0.2\pm0.3$ & $ 4.6\pm0.2\pm0.4$ \\
 0.9 & 39.7 & 5.0 & 15.9 & 5 & $4.8\pm0.2\pm0.4$ & $ 6.2\pm0.2\pm0.5$ \\
 1.6 & 50.3 & 5.0 & 22.1 & 5 & $5.6\pm0.1\pm0.6$ & $ 8.8\pm0.1\pm0.8$ \\
 2.5 & 60.2 & 2.1 & 30.2 & 5 & $7.6\pm0.1\pm1.0$ & $15.3\pm0.1\pm1.9$ \\
 \hline
 9.7 &  0.4 & 2.1 &  7.6 & 5 & $3.4\pm0.2\pm0.3$ & $ 3.3\pm0.2\pm0.2$ \\
 9.1 & 19.1 & 5.0 &  8.6 & 5 & $3.7\pm0.2\pm0.3$ & $ 3.9\pm0.2\pm0.3$ \\
 8.8 & 29.4 & 5.0 & 11.4 & 5 & $4.2\pm0.2\pm0.3$ & $ 4.8\pm0.2\pm0.4$ \\
 8.4 & 40.1 & 5.0 & 15.8 & 5 & $5.0\pm0.2\pm0.4$ & $ 6.5\pm0.2\pm0.5$ \\
 7.1 & 60.6 & 5.0 & 29.7 & 5 & $6.6\pm0.1\pm1.0$ & $13.2\pm0.1\pm1.9$ \\
 \hline
19.7 &  0.6 & 2.1 &  8.6 & 5 & $3.8\pm0.2\pm0.3$ & $ 3.6\pm0.2\pm0.3$ \\
19.3 & 20.1 & 5.0 & 10.2 & 5 & $4.3\pm0.2\pm0.3$ & $ 4.3\pm0.2\pm0.3$ \\
18.7 & 30.6 & 5.0 & 13.0 & 5 & $4.8\pm0.2\pm0.4$ & $ 5.2\pm0.2\pm0.4$ \\
18.3 & 41.5 & 5.0 & 16.6 & 5 & $5.5\pm0.2\pm0.4$ & $ 6.9\pm0.1\pm0.5$ \\
17.0 & 61.9 & 5.0 & 28.4 & 5 & $7.1\pm0.1\pm0.8$ & $14.3\pm0.1\pm1.6$ \\
\hline
29.7 &  0.9 & 2.1 &  9.8 & 5 & $4.6\pm0.2\pm0.3$ & $ 4.0\pm0.2\pm0.3$ \\
29.0 & 21.9 & 5.0 & 11.6 & 5 & $4.9\pm0.2\pm0.4$ & $ 4.6\pm0.2\pm0.3$ \\
28.6 & 32.8 & 5.0 & 14.5 & 5 & $5.6\pm0.2\pm0.4$ & $ 5.8\pm0.2\pm0.4$ \\
28.3 & 43.8 & 5.0 & 18.1 & 5 & $6.3\pm0.2\pm0.5$ & $ 7.6\pm0.1\pm0.6$ \\
26.9 & 63.7 & 5.0 & 30.1 & 5 & $7.8\pm0.1\pm0.9$ & $15.6\pm0.1\pm1.8$ \\
\hline
39.8 &  0.9 & 2.1 & 10.7 & 5 & $5.8\pm0.2\pm0.4$ & $ 4.4\pm0.2\pm0.3$ \\
38.9 & 24.5 & 5.0 & 12.9 & 5 & $6.0\pm0.2\pm0.4$ & $ 5.1\pm0.2\pm0.4$ \\
38.6 & 36.2 & 5.0 & 15.6 & 5 & $6.6\pm0.2\pm0.5$ & $ 6.4\pm0.2\pm0.5$ \\
38.2 & 47.3 & 5.0 & 19.4 & 5 & $7.5\pm0.1\pm0.6$ & $ 8.6\pm0.1\pm0.7$ \\
37.7 & 57.5 & 5.0 & 24.2 & 5 & $8.2\pm0.1\pm0.8$ & $12.1\pm0.1\pm1.1$ \\
36.9 & 66.3 & 5.0 & 31.6 & 5 & $8.9\pm0.1\pm1.1$ & $17.6\pm0.1\pm2.2$ \\
\end{tabular}
    \caption{Reconstructed telescope pointing resolution, which is the hit prediction precision on the DUT plane, as a function of the beam energy and incidence angles. A higher beam energy improves the pointing resolution (see first two rows), while for increasing incidence angles a larger distances between the telescope arms ($D^\mathrm{DUT}$) worsens the pointing resolution. $D^\mathrm{tel.}$ is the distance between the three telescope sensor planes within each arm.}
    \label{tab:hit_prediction_precision}
\end{table}

\paragraph{Intrinsic hit resolution}
The intrinsic spatial resolution is computed as the square root of the diagonals of $V^\mathrm{int}$ in Eq.~\ref{eq:residual_widths_2d}. Table~\ref{tab:depfet_resolution} summarises the obtained intrinsic hit resolutions at perpendicular incidence. The binary resolution $P/\sqrt{12}$ for a single pixel with pitch $P$ is indicated as well. At perpendicular incidence, mostly 1-pixel clusters with a small admixture of multi-pixel clusters are measured and the overall intrinsic resolution is slightly better than binary resolution. 

Table~\ref{tab:depfet_resolution_cog} gives a comparison of the intrinsic resolution between the centre of gravity and cluster shape position finding algorithms. The results are shown for incidence angles $\theta$ ranging from perpendicular incidence to around $60^\circ$, close to the maximum tracking acceptance of Belle~II. The cluster shape results are computed from a training sample with \SI{4E5}{clusters} per angle. A minimum number of 100 tracks per $\rho$ shape is required and hit positions covering on average \SI{98.8}{\percent} of all DUT clusters are computed. The number of distinct cluster shapes and shape corrections available depends strongly on the incidence angles with more shapes for larger angles. For perpendicular incidence, 25 shapes and 113 shape corrections are found, while for incidence angles of $\phi=30^\circ,\theta=60^\circ$, 84  shapes and 314 shape corrections are found.
It is observed that the spatial resolution does not further improve when increasing the number of $\rho$ bins beyond 14 for multi-pixel clusters. The spatial resolution provided by the cluster shape method is at least as good as the centre of gravity method, even in cases with intermediate incidence angles where the assumption of linear charge sharing is best met.

\begin{figure}[p]
    \centering
    \includegraphics[width=\textwidth,page=1]{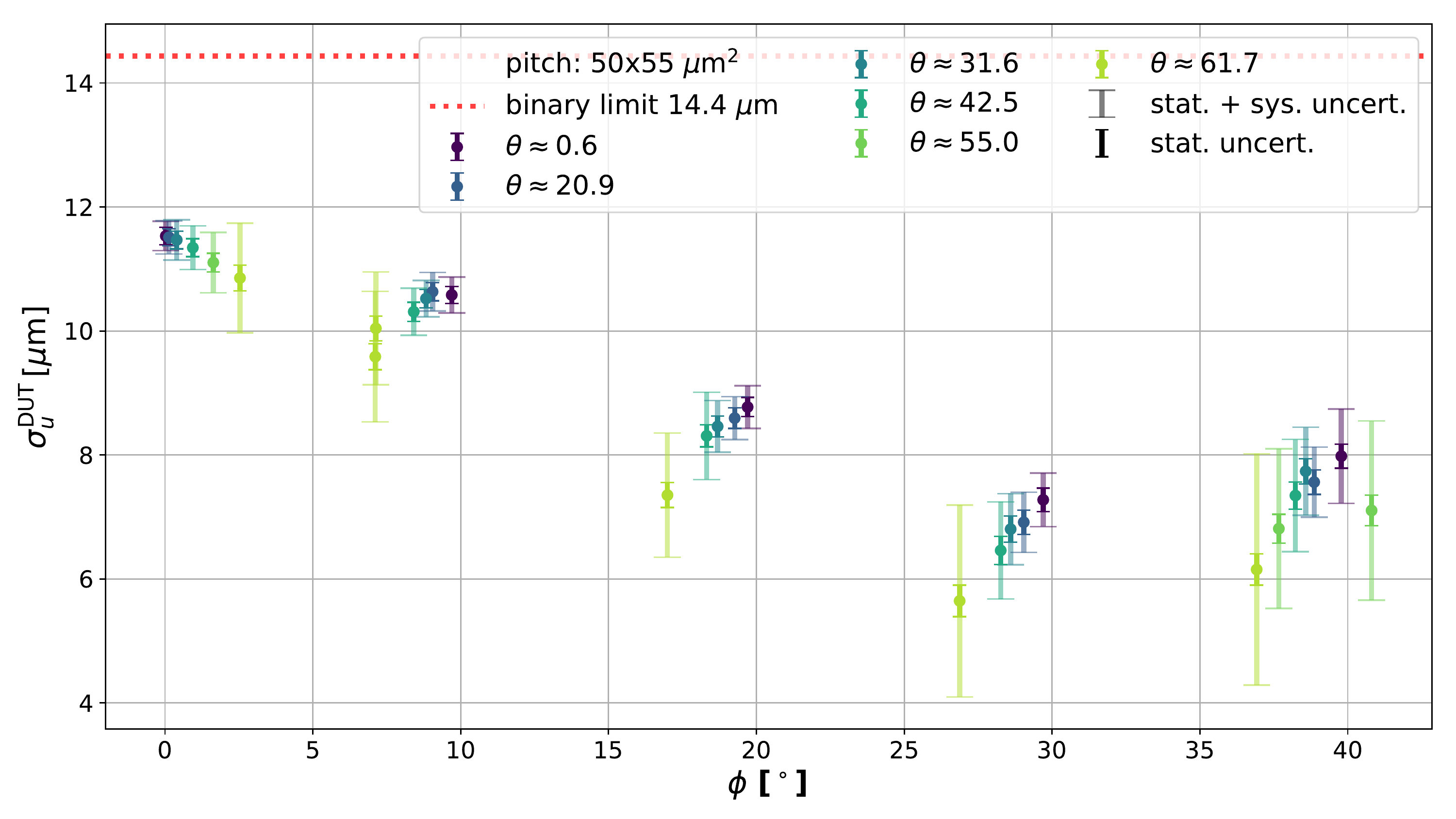}
    \includegraphics[width=\textwidth,page=6]{resolution_residuals.pdf}
    \caption{\textit{(colour online)} Intrinsic DUT hit resolutions, averaged over all cluster sizes, in $u$- and $v$-direction as a function of the respective incidence angle for a pixel pitch of $\SI{50}{\micro\metre}\times\SI{55}{\micro\metre}$.}
    \label{fig:depfet_resolution_angles}
\end{figure}

Figure~\ref{fig:depfet_resolution_angles} shows the DUT intrinsic hit resolution as a function of incidence angle. In $u$-direction (pitch \SI{50}{\micro\metre}), a minimum resolution of $\sigma^\mathrm{DUT}_u=\SI[parse-numbers=false]{5.4\pm0.3 (stat.)\pm1.3(syst.)}{\micro\metre}$
at $\phi=26.9^\circ$ is measured. In $v$-direction, a minimum resolution of $\sigma^\mathrm{DUT}_v=\SI[parse-numbers=false]{7.0\pm0.2 (stat.)\pm0.5(syst.)}{\micro\metre}$ at $\theta=36.2^\circ$ for a pitch of \SI{55}{\micro\metre} and $\sigma^\mathrm{DUT}_v=\SI[parse-numbers=false]{7.4\pm0.2 (stat.)\pm0.5(syst.)}{\micro\metre}$ at $\theta=36.2^\circ$ for a pitch of \SI{60}{\micro\metre} is measured. At these angles, the average overall cluster size is $2.8$ with about \SI{80}{\percent} of the clusters having a size of 2 in $u$- and $v$-direction. The best resolution is obtained for 2-column (2-row) clusters. The angle at which the optimal 2-column (2-row) cluster occurrence is maximised depends on the respective pixel pitch, as seen in the results.

\paragraph{Systematic uncertainty on resolution}
The statistical uncertainty of the spatial resolution is computed by the statistical uncertainty of the residual covariance matrix. There is an additional systematical uncertainty on the total residual RMS due to the residual drift. The average bias shift $\delta_b$ over consecutive runs is determined. Assuming this shift corresponds to an additional box distribution of width $\delta_b$ folded in, the systematical uncertainty on the residual width is estimated as $\delta_b/\sqrt{12}$, which is in the order of \SI{0.1}{\micro\metre}.

The systematic uncertainty on the spatial resolution is dominated by the systematic uncertainty on the telescope covariance matrix $T$, which is estimated by variation of the assumed beam energy by $\pm\SI{5}{\percent}$ and variation of the effective telescope sensor plane resolution by $\pm\SI{5}{\percent}$ used in the track fit. An additional systematic uncertainty component is the standard deviation of the distribution of track extrapolation uncertainties. The systematical uncertainty on the extracted DUT spatial resolution increases significantly for larger angles since the distance between the telescope arms needed to be increased to accommodate the tilted DUT housing. This worsens the telescope pointing resolution and increases its systematic uncertainty (cf.~Tab.~\ref{tab:hit_prediction_precision}). For large angles, $\sigma^\mathrm{DUT}$ and $\sigma^\mathrm{tel}$ thus become of comparable size and systematic effects have large impact.

\begin{table}[p]
\centering
\begin{tabular}{c|c|c|c}
 pitch $P_u$ & $P_u/\sqrt{12}$ & $\sigma_u^\mathrm{DUT}(L_u=1)$ & $\sigma_u^\mathrm{DUT}(L_u=2)$ \\
 \hline
 50 & 14.4 & $11.7\pm0.2\pm0.2$ & $5.7\pm0.2\pm0.2$\\
 \hline
 pitch $P_v$ & $P_v/\sqrt{12}$ & $\sigma_u^\mathrm{DUT}(L_v=1)$ & $\sigma_u^\mathrm{DUT}(L_v=2)$ \\
 \hline
 55 & 15.9 & $13.7\pm0.1\pm0.1$ & $6.9\pm0.2\pm0.2$\\
 60 & 17.3 & $15.2\pm0.1\pm0.1$ & $7.8\pm0.2\pm0.2$\\
 70 & 20.2 & $17.1\pm0.1\pm0.1$ & $8.3\pm0.2\pm0.2$\\
 85 & 24.5 & $20.9\pm0.1\pm0.1$ & $9.4\pm0.2\pm0.1$\\
\end{tabular}
\caption{Intrinsic DUT hit resolution per pixel pitch and column/row cluster size $L_{u/v}$ at perpendicular particle incidence for cluster shape based position finding. All values are given in units of \si{\micro\metre}. The uncertainties are given with the statistical and systematical contributions. An improvement over simple binary resolution for 1-pixel clusters is expected, since the exclusion of multi-pixel clusters reduces the effective pixel pitch.}
\label{tab:depfet_resolution}
\end{table}

\begin{table}[p]
\centering
\begin{tabular}{c|c|c|c}
  $\phi$ & $\theta$ & $^{\mathrm{cDB}}\sigma_v^\mathrm{DUT}\pm\mathrm{stat.}\pm\mathrm{syst.}$ & $^{\mathrm{CoG}}\sigma_v^\mathrm{DUT}\pm\mathrm{stat.}\pm\mathrm{syst.}$ \\
 \hline
 0.0 &     0.0 & $14.3\pm0.1\pm0.1$ & $14.6\pm0.1\pm0.1$ \\
 0.1 &    18.8 & $11.3\pm0.1\pm0.1$ & $11.8\pm0.1\pm0.1$ \\
 0.4 &    29.0 & $ 8.9\pm0.2\pm0.4$ & $ 9.5\pm0.2\pm0.3$ \\
 0.9 &    39.7 & $ 7.5\pm0.2\pm0.5$ & $ 7.8\pm0.2\pm0.4$ \\
 1.6 &    50.3 & $ 9.9\pm0.2\pm0.9$ & $10.5\pm0.2\pm0.8$ \\
 2.5 &    60.2 & $ 9.6\pm0.3\pm3.1$ & $11.0\pm0.3\pm2.7$ \\
 \hline
 9.7 &     0.4 & $14.3\pm0.1\pm0.1$ & $14.7\pm0.1\pm0.1$ \\
 9.1 &    19.1 & $11.1\pm0.1\pm0.2$ & $11.7\pm0.1\pm0.2$ \\
 8.8 &    29.4 & $ 8.7\pm0.2\pm0.2$ & $ 9.2\pm0.2\pm0.2$ \\
 8.4 &    40.1 & $ 7.6\pm0.2\pm0.5$ & $ 7.8\pm0.2\pm0.5$ \\
 7.1 &    60.6 & $10.2\pm0.3\pm3.3$ & $11.5\pm0.3\pm2.9$ \\
 \hline
19.7 &     0.6 & $14.4\pm0.1\pm0.1$ & $14.7\pm0.1\pm0.1$ \\
19.3 &    20.1 & $10.9\pm0.1\pm0.2$ & $11.5\pm0.1\pm0.2$ \\
18.7 &    30.6 & $ 8.5\pm0.2\pm0.3$ & $ 9.0\pm0.2\pm0.3$ \\
18.3 &    41.5 & $ 8.4\pm0.2\pm1.0$ & $ 8.4\pm0.2\pm1.0$ \\
17.0 &    61.9 & $11.5\pm0.2\pm2.1$ & $12.9\pm0.2\pm1.9$ \\
\hline
29.7 &     0.9 & $14.5\pm0.1\pm0.1$ & $14.9\pm0.1\pm0.1$ \\
29.0 &    21.9 & $10.2\pm0.1\pm0.2$ & $10.9\pm0.2\pm0.2$ \\
28.6 &    32.8 & $ 8.0\pm0.2\pm0.4$ & $ 8.4\pm0.2\pm0.4$ \\
28.3 &    43.8 & $ 8.6\pm0.2\pm0.7$ & $ 8.4\pm0.2\pm0.7$ \\
26.9 &    63.7 & $13.3\pm0.2\pm2.3$ & $14.3\pm0.2\pm2.2$ \\
\hline
39.8 &     0.9 & $14.4\pm0.1\pm0.2$ & $14.9\pm0.1\pm0.2$ \\
38.9 &    24.5 & $ 9.2\pm0.2\pm0.2$ & $10.0\pm0.2\pm0.2$ \\
38.6 &    36.2 & $ 7.4\pm0.2\pm0.5$ & $ 7.7\pm0.2\pm0.5$ \\
38.2 &    47.3 & $ 9.5\pm0.2\pm0.8$ & $ 9.2\pm0.2\pm0.8$ \\
37.7 &    57.5 & $10.5\pm0.2\pm1.6$ & $11.4\pm0.2\pm1.5$ \\
36.9 &    66.3 & $15.5\pm0.2\pm2.6$ & $16.5\pm0.2\pm2.5$ \\
\end{tabular}
\caption{Comparison of the intrinsic spatial resolution when using the cluster shape based (cDB) and the CoG based PFA, based on the \SI{60}{\micro\metre} pitch pixels. In general, the cluster shape based PFA yields a better spatial resolution. Note that the systematical errors are fully correlated between cDB and CoG method.}
\label{tab:depfet_resolution_cog}
\end{table}

\FloatBarrier

\section{DUT response to particle beam}
\label{sec:dut_performance}
\begin{figure}[p]
  \centering
  \includegraphics[width=\textwidth]{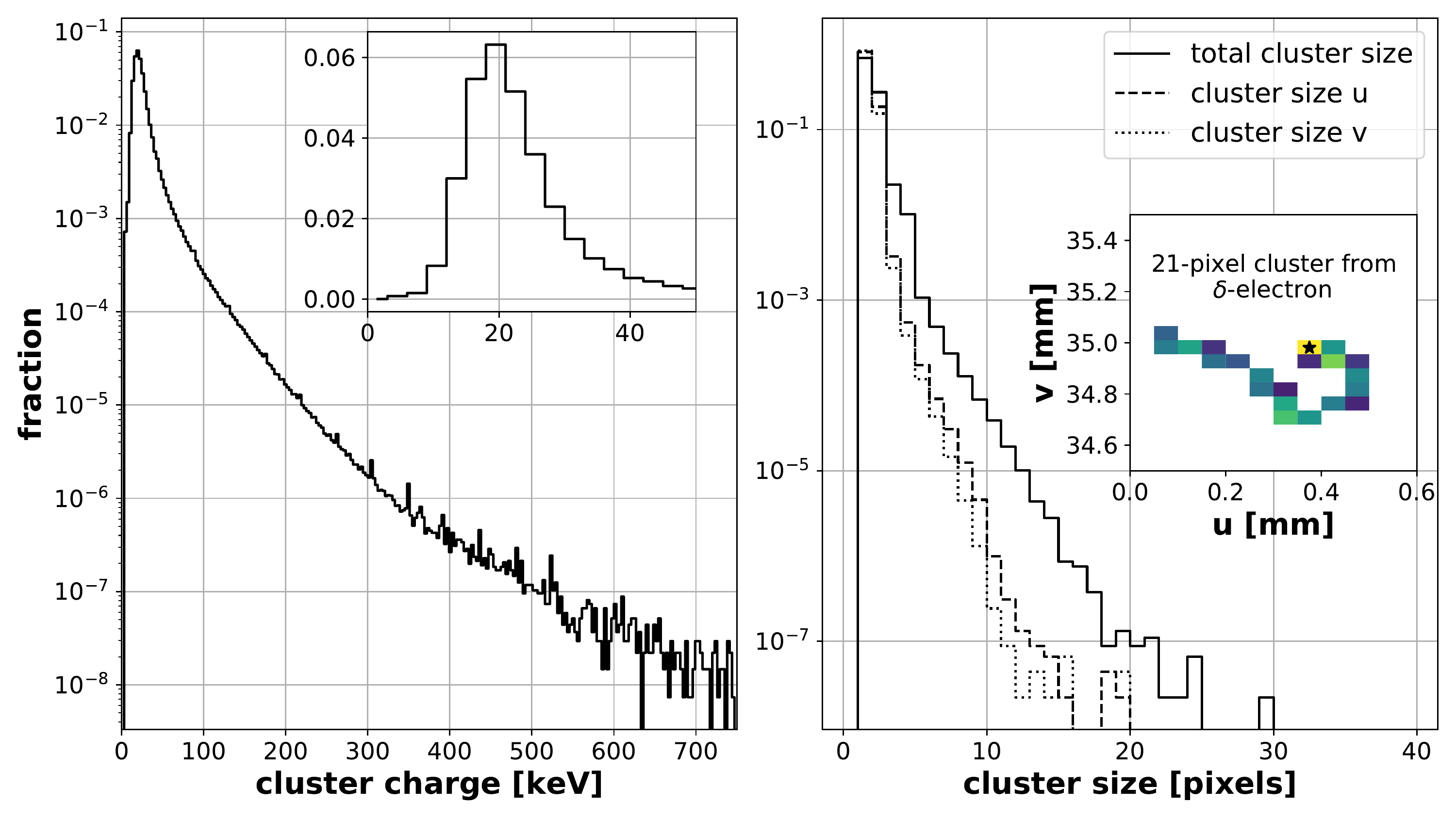}
  \caption{\textit{(colour online)} \textit{left:} Typical cluster charge distribution for \SI{3}{\giga\electronvolt} electrons for DUT \WfourtyIF\ at perpendicular incidence. The lowest cluster charge signal is \SI{5}{ADU} since a zero-suppression threshold of $>\SI{4}{ADU}$ is used. The Landau most probable value is \SI{19.7}{\kilo\electronvolt}. The $y$-axis is bin counts normalised to unity. The highest bin is not an overflow bin. \textit{right:} Corresponding cluster size distribution. Only clusters matched to reconstructed tracks are considered. About \SI{96}{\percent} of all cluster are 1- or 2-pixel clusters. The average total cluster size is $1.36$. The $y$-axis is bin counts normalised to unity. The highest bin is not an overflow bin. Clusters with sizes $>20$ are occasionally found and are due to $\delta$-electrons. An example of such a cluster is shown with absolute sensor coordinates and pixel signals indicated by the colour (yellow: large signal, blue: small signal) and the extrapolated track intersection indicated by a star.}
  \label{fig:cluster_charge_and_size_distribution}
\end{figure}

The amount of charge generated by charged particles traversing a thin silicon sensor is distributed according to a Landau distribution~\cite{PhysRevD.98.030001}. The peak of the distribution gives the most probable charge deposition, $Q_\mathrm{MPV}$. In order to calibrate the DUT signals to absolute energies, a Geant4~\cite{Geant42003} simulation is employed that simulates the energy deposition of \SI{3}{\giga\electronvolt} electrons in \SI{75}{\micro\metre} silicon. A Landau distribution is fitted to the simulated data to extract the expected $\hat{Q}_\mathrm{MPV}$ in units of \si{\kilo\electronvolt}. A Landau distribution folded with a Gaussian distribution (to account for detector noise) is fitted to the measured cluster charge distribution to extract the measured $Q_\mathrm{MPV}$ in ADU. The calibration factor $k$ is obtained as the ratio $k=\hat{Q}_\mathrm{MPV}/Q_\mathrm{MPV}$ and is in the order of \SI{0.8}{\kilo\electronvolt/ADU} (Tab.~\ref{tab:dut_signal_response}). A gradient in cluster charge MPV along the $v$-direction (columns) in the order of \SI{20}{\percent} is observed, with pixels closer to the ASICs having lower $Q_\mathrm{MPV}$ than pixels at the far end of the modules. Thus, the conversion factors are determined individually per pixel row. When measured signals or thresholds are given in number of electrons, values in \si{\kilo\electronvolt} were divided by \SI{3.6}{\electronvolt} which is the energy necessary to create an electron-hole pair in silicon. Figure~\ref{fig:cluster_charge_and_size_distribution} shows typical cluster charge and cluster size distributions for perpendicular incidence. Table~\ref{tab:dut_signal_response} summarises the overall cluster charge MPV, calibration factors (average and standard deviation over the sensor rows), signal-to-noise ratios (S/N), thresholds and fraction of alive (non-masked) pixels for the measured DUTs. The signal-to-noise ratio (S/N) is determined as the ratio of $Q_\mathrm{MPV}$ over the average pixel noise (cf. Tab.~\ref{tab:pxd_sensor_modules}).

\begin{table}[p]
\centering
\begin{tabular}{l|c|c|c|c|c}
 DUT & $Q_\mathrm{MPV}$ [ADU] & $k$ [\si{\kilo\electronvolt/ADU}] & S/N & thrsd. [$e^-$] & alive [\si{\percent}] \\
 \hline
\WfourtyIF       & $20.3\pm0.5$ & 1.01 (0.33) & $25\pm6$ & $1020-1770$ (20) & 99.1 \\
\WelevenOFtwo    & $29.7\pm0.5$ & 0.68 (0.12) & $37\pm9$ & $740-1150$ (15) & 97.4 \\
\WfiveOBone      & $26.6\pm0.5$ & 0.74 (0.07) & $44\pm15$ & $850-1290$ (16) & 74.2 \\
\WfiveOBoneIrrad & $30.2\pm0.5$ & 0.69 (0.23) & $50\pm17$ & $960-1250$ (62) & 73.1 \\
\WphaseTwo       & $40.1\pm0.5$ & 0.51 (0.12) & $57\pm16$ & $520-1930$ (11) & 99.8 \\
\end{tabular}
\caption{Cluster signal response of the DUTs to a \SI{3}{\giga\electronvolt} electron beam at perpendicular incidence. $Q_\mathrm{MPV}$ is determined from the cluster signal distribution over the full sensor. The variations in the MPV among modules originates from different tuning of the DEPFET gate potentials and digitisation gains (cf.~Tab.~\ref{tab:pxd_sensor_modules}). The calibration factors $k$ are given as the average over the full sensor and the standard deviation in parentheses. The spread in S/N among the DUTs is due to different operation points and pedestal spreads (cf. Fig.~\ref{fig:pedestal_distribution} and Tab.~\ref{tab:pxd_sensor_modules}). The thresholds are given in number of electrons as the 1-to-99 percentile range over all pixels and corresponds to the $>\SI{4}{ADU}$ online pixel signal threshold ($>\SI{5}{ADU}$ for \WfiveOBoneIrrad). The uncertainty on the thresholds is given by the uncertainty on the gain $k$ and is indicated in brackets. The fraction of alive pixels is the number of pixels that show a non-zero signal response over all $250 \times 768$ pixels. The modules \WelevenOFtwo\ and \WfiveOBone\ are grade-B with $<\SI{99}{\percent}$ alive pixel. The module \WfiveOBone\ has one of four ASIC pairs not functioning.}
\label{tab:dut_signal_response}
\end{table}

\begin{figure}[p]
    \centering
    \includegraphics[width=\textwidth]{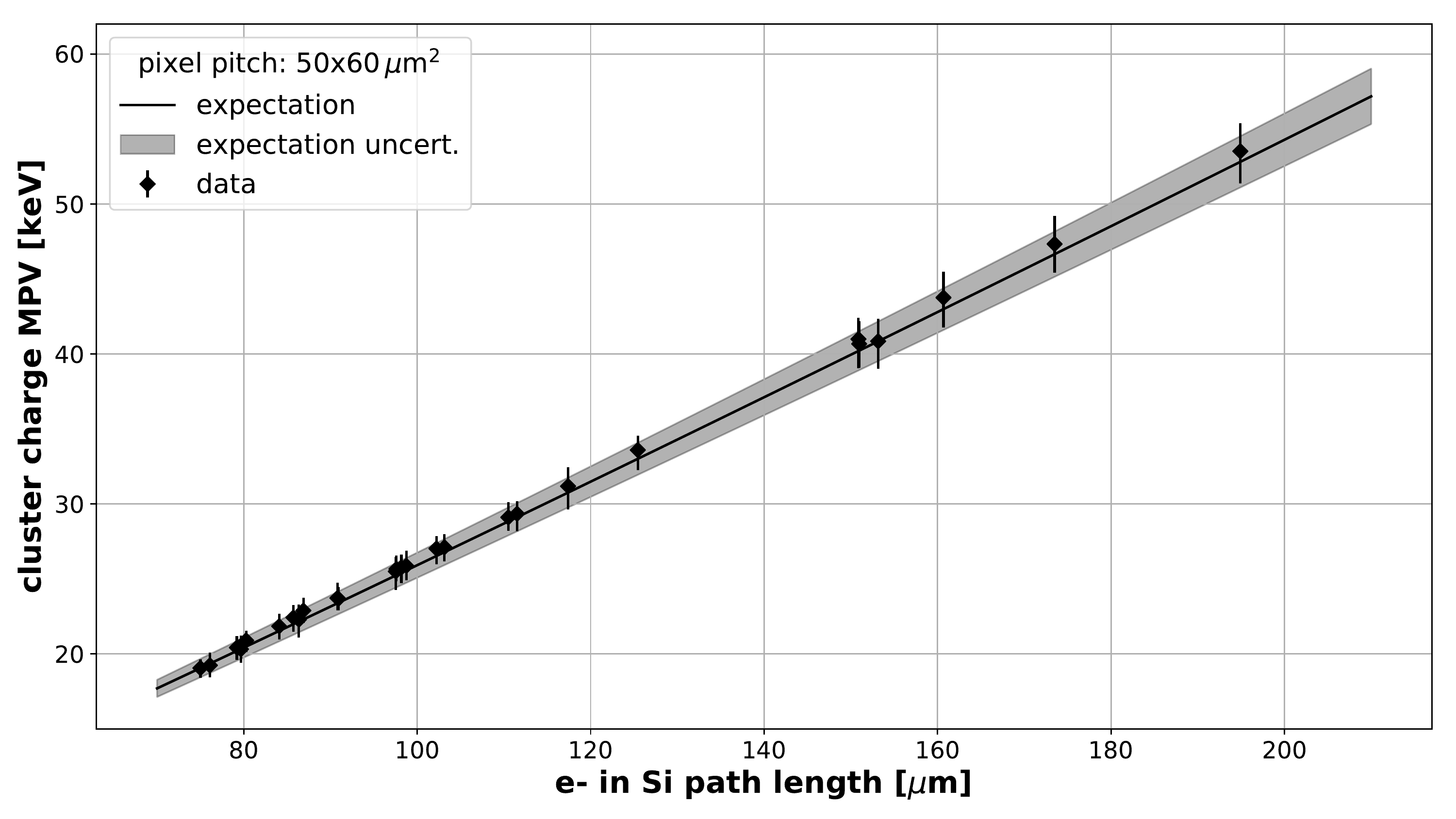}
    \caption{The most probable value $Q_\mathrm{MPV}$ of the deposited cluster charge as a function of the path length of the electron track in the sensitive volume of the \SI{75}{\micro\metre} thick sensor. The data points indicate the median and 25-to-75-percentile range of all super-pixels $Q_\mathrm{MPV}$ and were recorded at different incidence angles resulting in an increased path length. The data points are compared with expected values. The expectation uncertainty is given by the uncertainty of the reference point at perpendicular incidence.}
    \label{fig:mpv_vs_path_length}
\end{figure}

\begin{figure}[p]
    \centering
    \includegraphics[width=\textwidth]{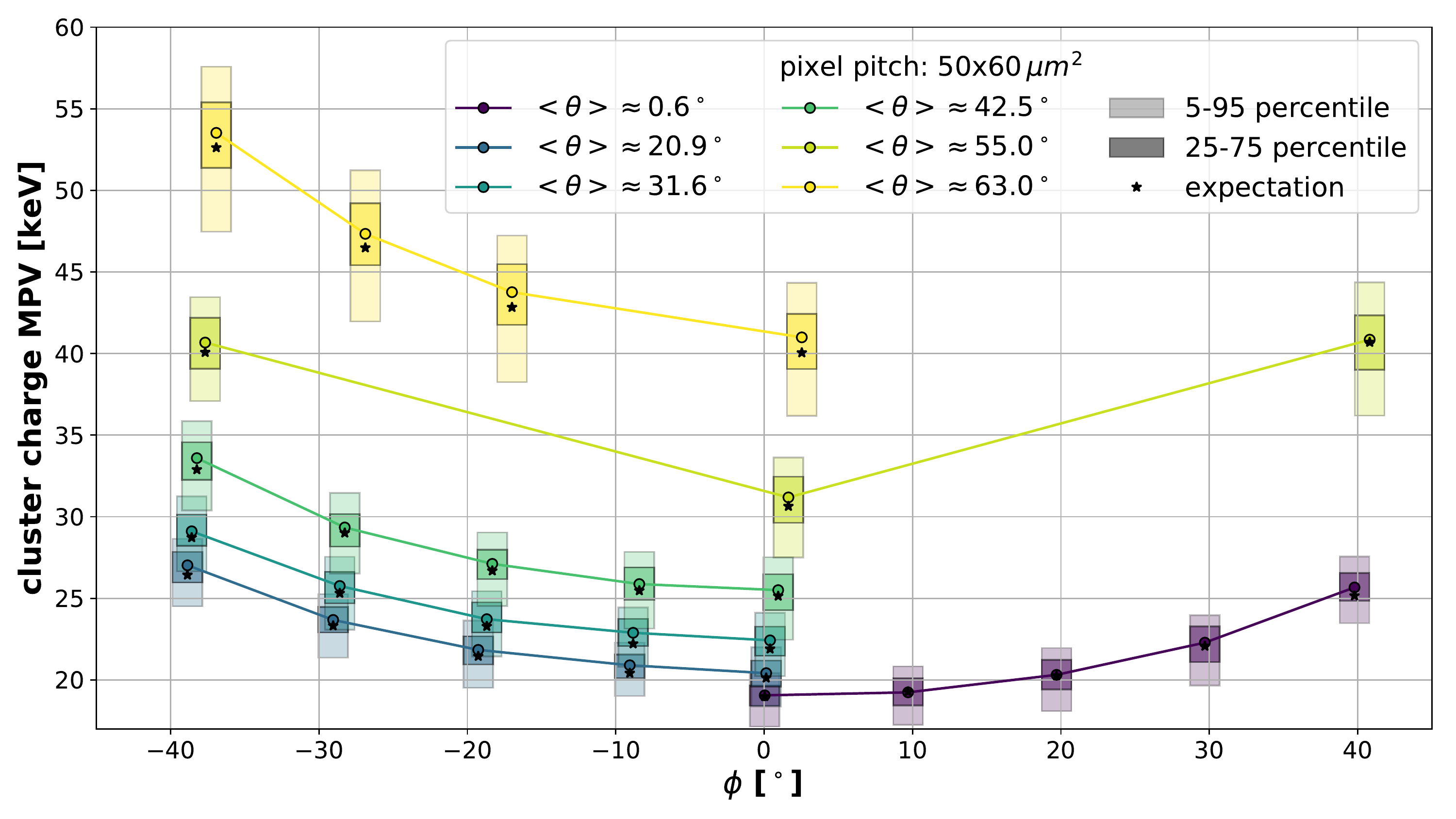}
    \caption{\textit{(colour online)} The most probable value $Q_\mathrm{MPV}$ of the deposited cluster charge as a function of the incidence angles $\phi$ and $<\theta>$ for a pixel pitch of $\SI{50}{\micro\metre}\times\SI{60}{\micro\metre}$. For each combination of angles, the coloured points indicate the median cluster charge $Q_\text{MPV}$ over all super-pixels bins (see text). The coloured bars indicate the respective 5-to-95 and 25-to-75 percentile ranges of the cluster charge MPVs. The results match the expectation (black stars) with a small but systematic overshoot for large angles, and confirm the symmetry in $\phi$.}
    \label{fig:mpv_vs_angle}
\end{figure}

\begin{figure}[p]
    \centering
    \includegraphics[width=\textwidth]{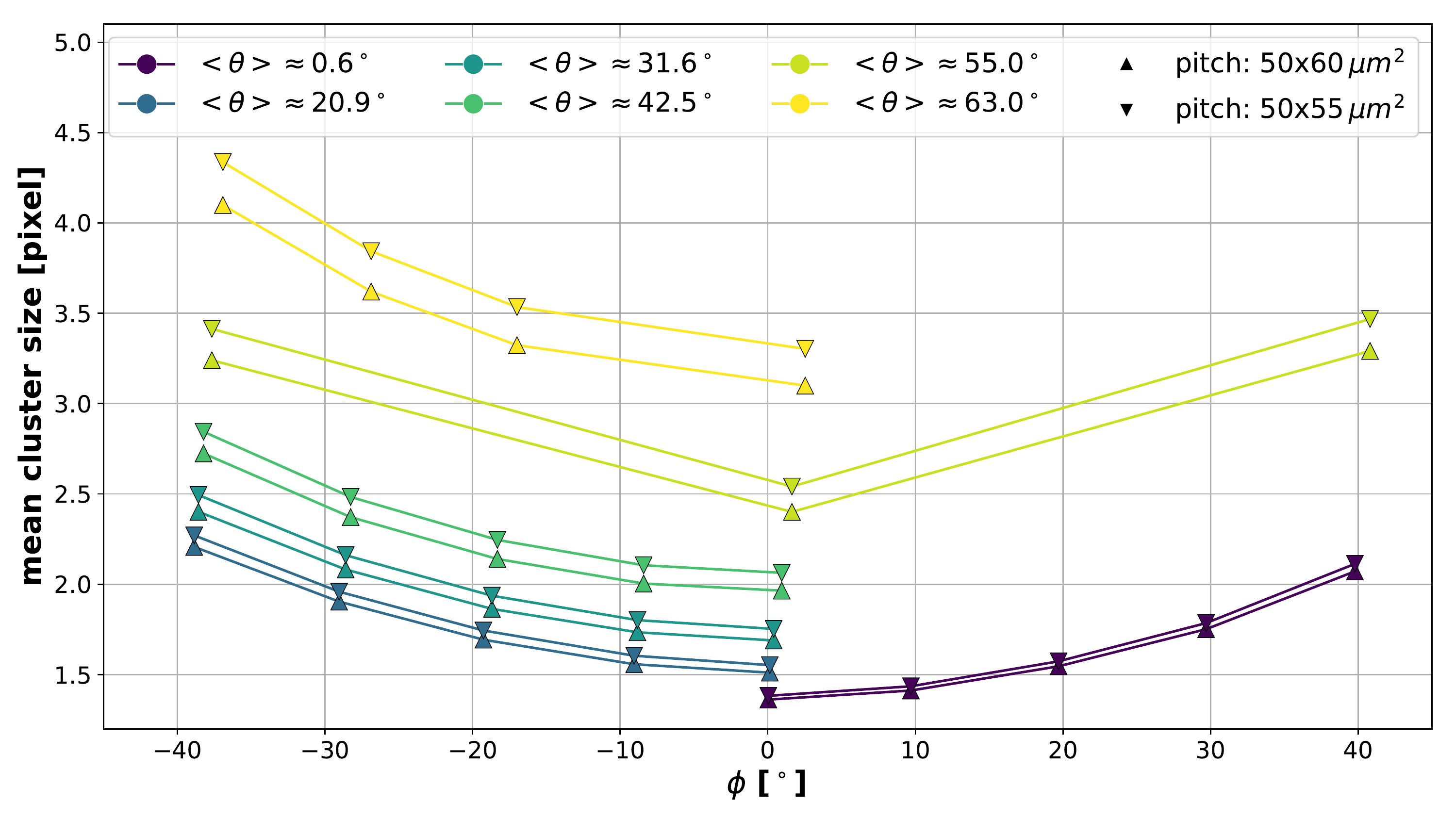}
    \caption{\textit{(colour online)} Average total size of clusters created by particles with inclined incidence on the DUT's pixel matrix as a function of the angles $\phi$ and $<\theta>$, separately for the small and large pixel pitch. The results meet the expectation that for larger angles the average cluster size grows. The difference in mean cluster size between pixels with small pitch compared to pixels with large pitch increases for larger angles.}
    \label{fig:cluster_size_vs_angle}
\end{figure}

\paragraph{Inclined particle incidence and charge sharing}
For inclined particle incidence, the charged particle travels a longer distance in the sensitive volume and thus deposits more energy and generates more charge compared to perpendicular incidence. The most probable value of the deposited cluster charge, $Q_\mathrm{MPV}$, increases with the path length of the electrons in the sensitive volume. From basic geometrical considerations, the total path length depends on the incidence angles as $l(\phi,\theta)=l_0 \sqrt{1 + \tan^2{\phi} + \tan^2{\theta}}$, where $l_0=\SI{75}{\micro\metre}$ is the path length at perpendicular incidence. The most probable energy loss per distance also scales with the total path length such that the expected cluster charge MPV as a function of the angles is
\begin{equation}
    \hat{Q}_\mathrm{MPV}(\phi,\theta) = Q_\mathrm{MPV}(0,0) \frac{\Delta_p(l(\phi,\theta), \beta, \gamma)}{\Delta_p(l_0, \beta, \gamma)} \ ,
    \label{eq:mpv_extrapolation}
\end{equation}
where $\beta$ and $\gamma$ are the relativistic parameters of the impinging electron, $Q_\mathrm{MPV}{(0,0)}$ is the reference value at perpendicular incidence and the material-dependent function $\Delta_p$ is given in Eq.~33.12 in~\cite{PhysRevD.98.030001}. Values for $Q_\mathrm{MPV}$ are determined as a function of the position on the sensor (pixel coordinates). Since single pixels do not collect a sufficient number of hits for reliably fitting a Landau distribution, groups of $u\times v=10\times8$ pixels, referred to as super-pixels, are formed. The average $Q_\mathrm{MPV}{(0,0)}$ over all super-pixels at perpendicular incidence is used as the reference point with an uncertainty given by the standard deviation over all super-pixels. The expected $\hat{Q}_\mathrm{MPV}(\phi,\theta)$ at angles $\phi,\theta$ is computed using Eq.~\ref{eq:mpv_extrapolation}. The uncertainty of the expectation is given by propagating the uncertainty of $Q_\mathrm{MPV}{(0,0)}$ in Eq.~\ref{eq:mpv_extrapolation}. At each angle, the expectation is compared with the averaged $Q_\mathrm{MPV}(\phi,\theta)$ over all super-pixels with an uncertainty given by the standard deviation over all super-pixels. Only clusters matched to a reconstructed track are used for determining $Q_\text{MPV}$. Figure~\ref{fig:mpv_vs_path_length} shows a good agreement of the measured and expected values of $Q_\mathrm{MPV}(\phi,\theta)$ at different path lengths.
The comparison can also be done for the individual angles as shown in Figure~\ref{fig:mpv_vs_angle}.
Electrons traversing the sensitive volume with an inclined path should produce larger clusters since the generated charge is spread over a larger area. This is confirmed in Fig.~\ref{fig:cluster_size_vs_angle}, which shows the average total cluster size as a function of the incidence angles. The data points are split into two groups corresponding to clusters formed by pixels of only one of the two pixel pitches. As expected, the smaller pixel pitch results in larger clusters on average.

\FloatBarrier

\section{Hit efficiency studies}
\label{sec:efficiency}
The hit efficiency of a sensor for a vertex detector is an important performance criterion. It measures the probability for the sensor to produce a hit signal above the zero-suppression threshold for a traversing charged particle. The hit efficiency can be measured in a beam test using a reference telescope by calculating the probability for a reconstructed particle track to have an associated hit on the DUT. The hit efficiencies of all available DUTs were measured.
The beam spot size ($2\times1\,\si{\centi\metre\squared}$) was smaller than the DUT's sensitive area ($6.1(4.5)\times1.2\,\si{\centi\metre\squared}$). The DUT was placed on movable stages allowing for movements within the plane perpendicular to the electron beam. The full sensor area was then measured in several ``illumination windows''. For each window, data were collected over equally long time periods and later combined. For the DUT \WfiveOBone\, before and after irradiation, only half of the sensor (ASIC pairs 1 and 2, columns 0-125) was illuminated and considered in the following analysis.

The hit efficiency is determined as a function of the extrapolated track impact position on the DUT in pixel coordinates. The binning of impact coordinates is chosen such that each bin collects a sufficient number of tracks, allowing for the determination of the efficiency with low statistical uncertainty, while ensuring a uniform distribution of tracks over the bin. With a binning accumulating $u\times v = 10\times8$ pixels into one bin (super-pixel), the average number of reconstructed tracks per super-pixel is $10^4$. A fraction of \SI{96}{\percent} of these super-pixel bins collect more than 4000 tracks. The remaining \SI{4}{\percent} of pixels collect on average 2600 tracks with a minimum of 565 tracks. 

\paragraph{Track selection and hit efficiency calculation}
The selection of reconstructed tracks, as they come out of the last step of the event reconstruction, was tightened and a hit on the FE-I4 plane was required for hit efficiency studies. Moreover, tracks with a predicted intersection at the DUT plane that is in the area of a pixel that is considered non-functional (dead or hot) are rejected. Given a set of $n$ tracks, the hit efficiency $\epsilon$ is calculated as the ratio of the number of tracks which have an associated cluster hit on the DUT $n_\mathrm{matched}$ and the total number of tracks $n_\mathrm{total}$:
\begin{equation}
    \epsilon = \frac{n_\mathrm{matched}}{n_\mathrm{total}} \ .
    \label{eq:efficiency}
\end{equation}
The uncertainty $\sigma_\epsilon$ on the hit efficiency is given by the variance of a binomial distribution:
\begin{equation}
    \sigma_\epsilon = \sqrt{\frac{\epsilon(1-\epsilon)}{n_\mathrm{total}}} \ .
    \label{eq:efficiency_error}
\end{equation}

\begin{figure}[p]
  \centering
  \includegraphics[width=\textwidth]{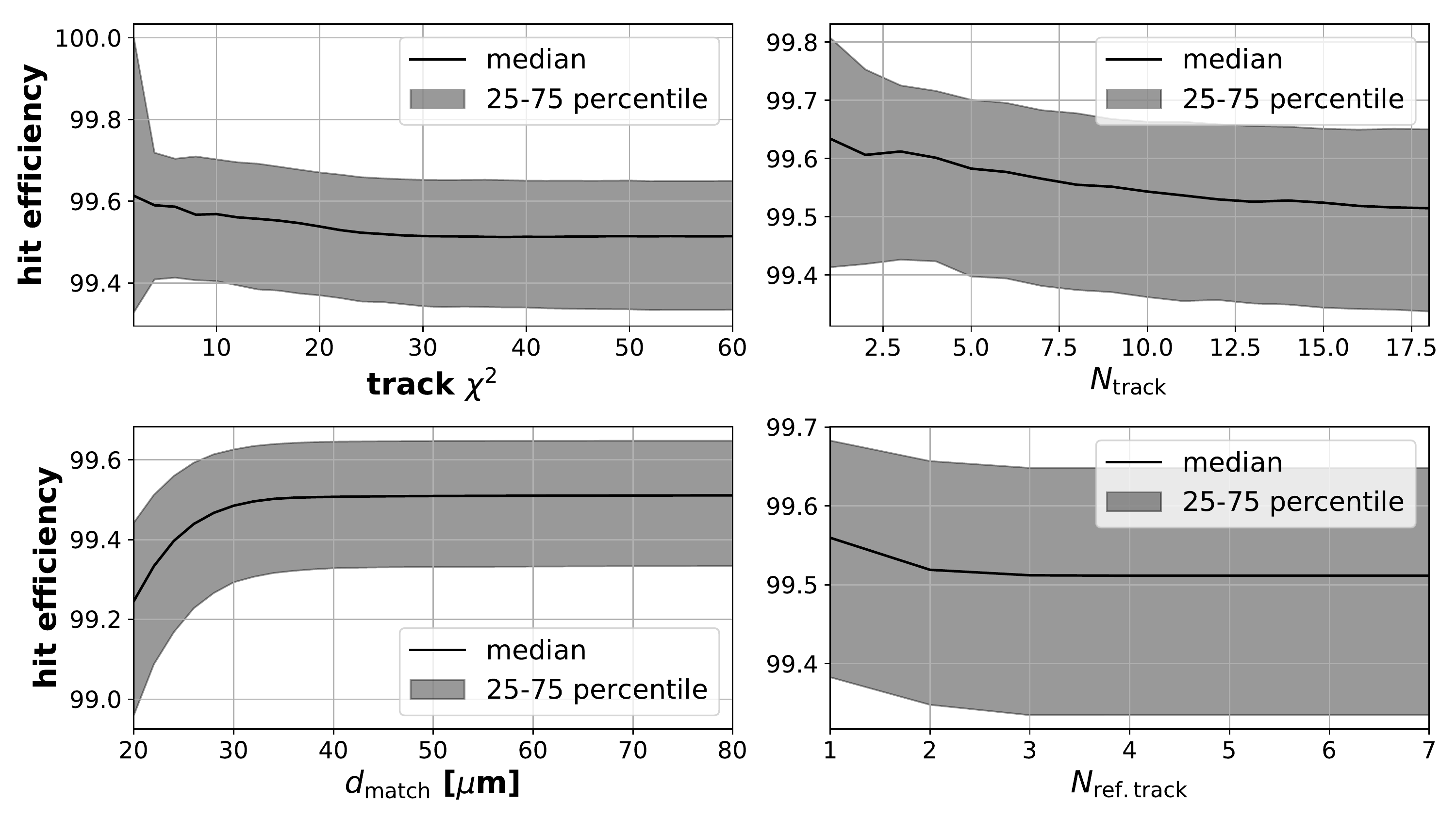}
  \caption{Hit efficiency as a function of several event and track properties indicating the systematic uncertainty due to fake tracks. Shown are, from left-to-right, top-down the maximum track $\chi^2$ value, the number of reconstructed tracks per event, the cluster hit-to-track matching distance in \si{\micro\metre} and the number of reconstructed tracks with a hit on the reference timing plane. The efficiency is given as the median and the 25-to-75 percentile range over all super-pixel efficiencies.}
  \label{fig:efficiency_systematics}
\end{figure}

\paragraph{Systematic uncertainty on hit efficiency}
The efficiency calculation should be robust with respect to the selection of reconstructed tracks. Influences of track selection criteria on the computed efficiency numbers should be evaluated to confirm, that the track reconstruction is under good control. Here, the systematic uncertainty on the hit efficiency is estimated based on the influence of event and track properties that potentially increase the probability of misreconstructed tracks. Misreconstructed tracks, for example due to including noise hits or due to ambiguities in assigning hits from multiple tracks in an event, cannot be matched to hits on the DUT, causing an artificial inefficiency. Such event properties are the track multiplicity, $N_\text{track}$, (number of reconstructed tracks per event) and the reference track multiplicity, $N_\text{ref. track}$, (number of reconstructed tracks with a hit on the reference FE-I4 timing plane per event). Additionally, the track $\chi^2$ value is considered. Figure~\ref{fig:efficiency_systematics} shows the hit efficiency as a function of thresholds applied to the respective properties. A small increase in the average hit efficiency is observed for $N_\text{track}\leq5$, $N_\text{ref. track}=1$ and $\chi^2\leq10$, respectively. The overall systematic uncertainty is determined as the difference of the averaged efficiency for the above event selection and for no additional event selection criteria. This procedure very consistently yields a systematic uncertainty of $\sigma_\epsilon^\mathrm{syst.}=\SI{0.1}{\percent}$ for all tested sensor modules. It is shown that the investigated dependencies are uniform and thus small variations in the track selection cuts do not significantly change the obtained efficiencies.
The systematic influence distance threshold, $d_\mathrm{match}$, of the hit-to-track matching is found to be negligible for values $d_\mathrm{match}\geq\SI{35}{\micro\metre}$. This confirms that the DUT alignment is well under control.

\begin{figure}[p]
    \centering
    \includegraphics[width=\textwidth]{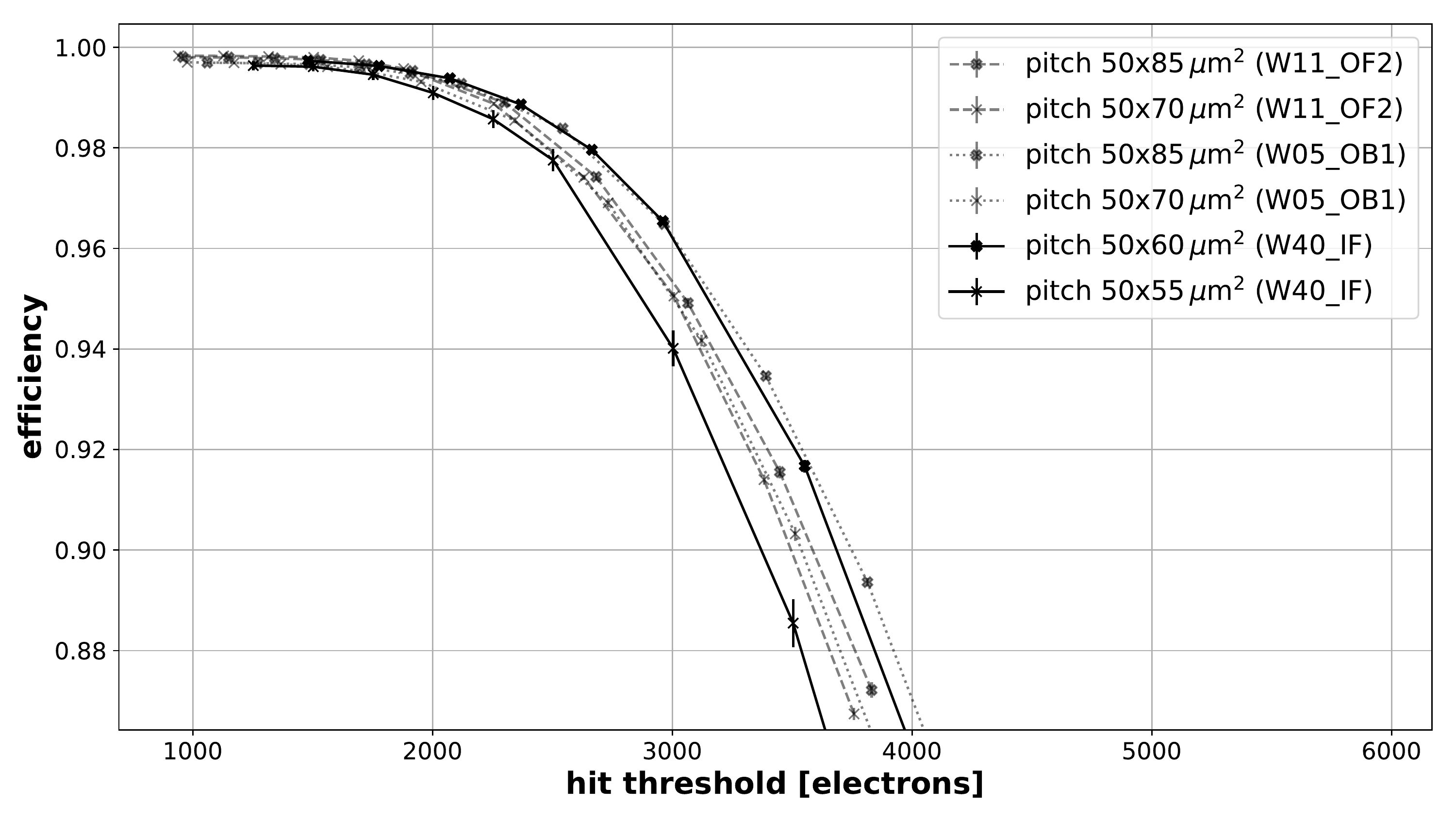}
    \caption{Hit efficiency as a function of the applied zero-suppression threshold in number of electrons for three of the measured DUTs separated small and large pixel pitch. The hit efficiency drops below \SI{99}{\percent} for thresholds above $2000$ electrons.}
    \label{fig:threshold_scan}
\end{figure}

The influence of the zero-suppression threshold on the hit efficiency was studied. Data recorded with an online zero-suppression thresholds of $>\SI{4}{ADU}$ (cf.~Tab.~\ref{tab:dut_signal_response}) were reprocessed with an increased \textit{offline} zero-suppression threshold. Figure~\ref{fig:threshold_scan} shows the degradation of hit efficiencies for increasing thresholds for different DUTs and pixel pitches. The respective lowest threshold point corresponds to the $>\SI{4}{ADU}$ online threshold, translating to 900 to 1400 electrons, depending on the respective DUT. The average hit efficiency over the sensor drops below \SI{99}{\percent} for signal thresholds above $2000$ electrons.

\begin{figure}[p]
    \centering
    \includegraphics[width=\textwidth]{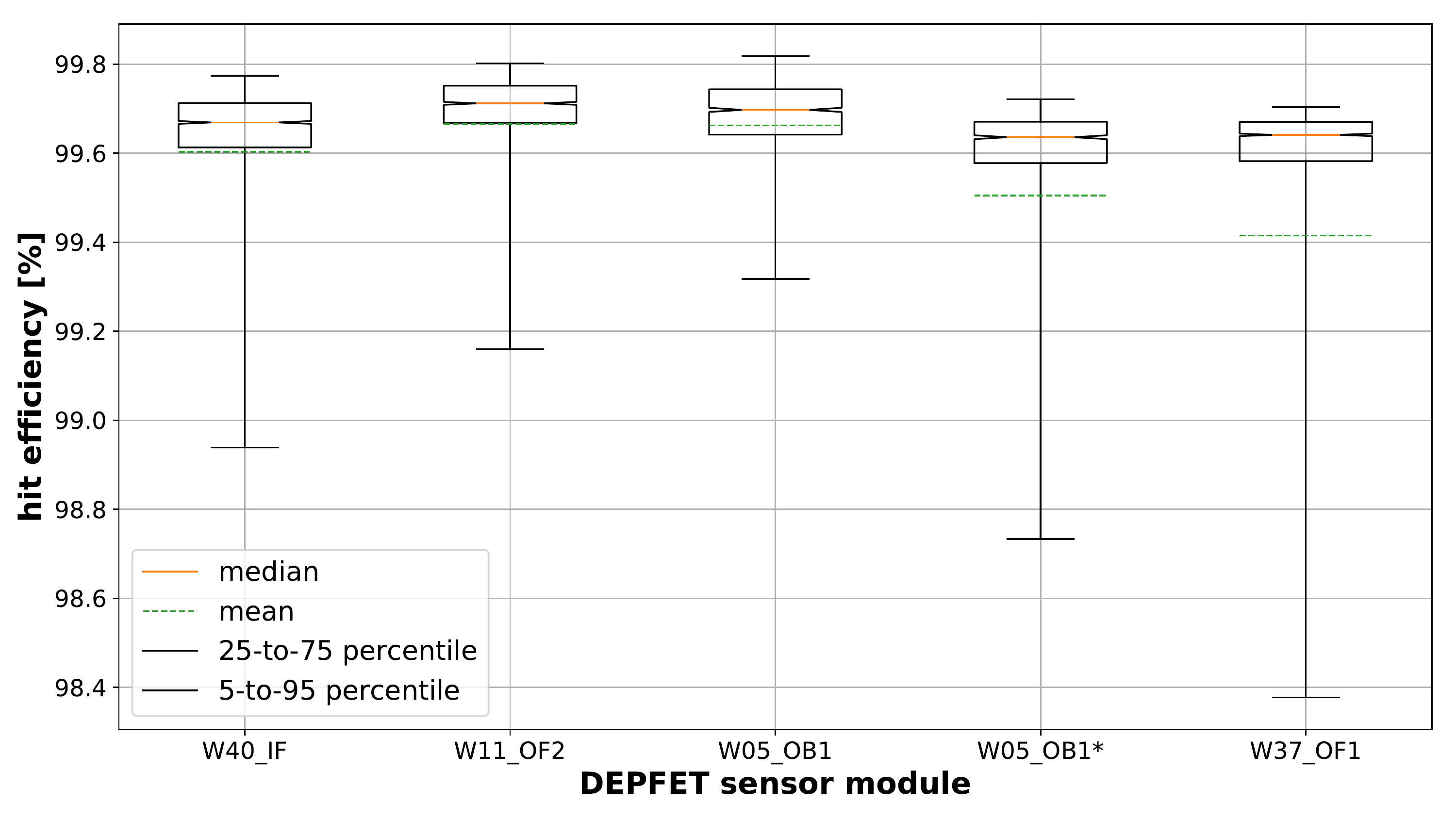}
    \caption{\textit{(colour online)} Comparison of the distribution of hit efficiencies over all super-pixels per DUT. The box gives the inter-quartile range (25-to-75 percentile range), the orange line within the box indicates the median and the dashed green line indicates the average over hit efficiencies in all super-pixels. The outer whiskers give the 5-to-95 percentile range of the distribution.}
    \label{fig:efficiency_comparison}
\end{figure}

\paragraph{Spatially resolved hit efficiency}
Assuming an efficiency of \SI{99}{\percent}, the $10^4$ collected tracks per super-pixel correspond to an expected statistical uncertainty of $\sigma_\epsilon=\SI{0.1}{\percent}$ per super-pixel. Table~\ref{tab:efficiencies} summarises the hit efficiencies of the tested DUTs averaged over all super-pixels. Figure~\ref{fig:efficiency_comparison} provides more details on the distribution of hit efficiencies over the super-pixels. All DUTs' hit efficiencies are in good agreement with the overall mean of \SI{99.6}{\percent}. The fraction of hot pixels is less than $\SI{1}{\percent}$ for all modules except the irradiated \WfiveOBoneIrrad. It shows overall higher fractions of dead and hot pixels due to an operator mistake during the irradiation campaign where few read-out channels were damaged by applying wrong potentials.

\begin{table}[p]
\centering
\begin{tabular}{l|c|c|c|c}
 & $\bar\epsilon\pm\text{stat.}\pm\text{syst.}$ {[}\%{]} & hot pixels {[}\%{]} & dead pixels {[}\%{]} & thrsd. [$e^-$]\\
 \hline
\WfourtyIF       & $99.60 \pm 0.06 \pm 0.1$ & 0.74 & 1.59 & $1300\pm140$ \\
\WelevenOFtwo    & $99.66 \pm 0.04 \pm 0.1$ & 0.67 & 4.40 & $940\pm100$ \\
\WfiveOBone      & $99.66 \pm 0.08 \pm 0.1$ & 0.09 & 0.44 & $1030\pm100$ \\
\WfiveOBoneIrrad & $99.50 \pm 0.02 \pm 0.1$ & 3.72 & 4.58 & $1030\pm60$ \\
\WphaseTwo       & $99.42 \pm 0.03 \pm 0.1$ & 0.95 & 0.62 & $710\pm 170$\\
\end{tabular}
\caption{Hit efficiency for all tested DUTs averaged over all super-pixels. The statistical uncertainty is the average statistical uncertainty of all super-pixels per sensor. The hot and dead pixel fractions are the number of masked hot and dead pixels over all $250\times768$ DEPFET pixels. A threshold of $>\SI{4}{ADU}$ was used for all DUTs, except for the irradiated \WfiveOBoneIrrad\ which used $>\SI{5}{ADU}$, translating to an averaged threshold in electrons (with standard deviation) as given in the right-most column (cf. Tab.~\ref{tab:dut_signal_response}).}
\label{tab:efficiencies}
\end{table}

\begin{figure}[p]
    \centering
    \includegraphics[width=\textwidth]{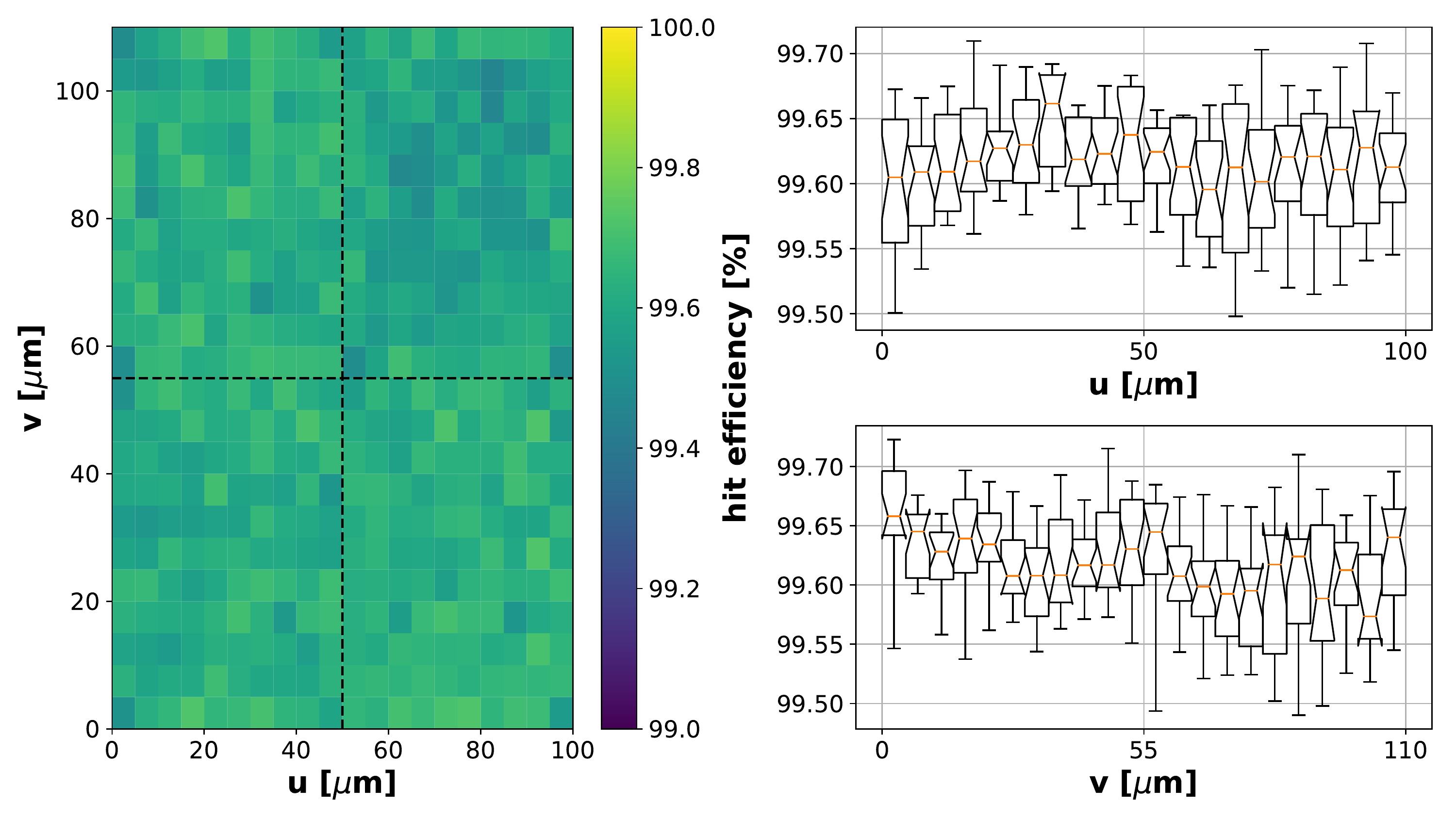}
    \caption{\textit{(colour online)} Sub-pixel spatially resolved hit efficiency within the $2\times2$ DEPFET pixel layout at perpendicular particle incidence. The box plots of the projections in $u$ and $v$ follow the definitions given in Figure~\ref{fig:efficiency_comparison}. The efficiency across the four-pixel layout is homogeneous, also at the borders between pixels.}
    \label{fig:efficiency_inpix}
\end{figure}

\paragraph{Sub-pixel hit efficiency}
The EUDET beam telescope provides a very good pointing resolution of about \SI{3.7}{\micro\metre} for appropriate beam energies and telescope geometries (cf.~Tab.~\ref{tab:hit_prediction_precision}). This is smaller than the DEPFET pixel pitches used in the PXD. For perpendicular incidence at \SI{3}{\giga\electronvolt}, a sub-pixel analysis of the hit efficiency was performed. A $2\times2$ grid of pixels (unit cell) is considered for the two regions of pixel pitches. Due to a limited number of hits per pixel, all hits are overlaid in a generic unit cell. For each reconstructed track, the extrapolated hit position on the DUT plane is converted to a relative hit position within the generic unit cell. These $u$- and $v$-coordinates are binned in $\SI{5}{\micro\metre}\times\SI{5}{\micro\metre}$ sized bins. Figure~\ref{fig:efficiency_inpix} shows the sub-pixel hit efficiency map and the projections in $u$- and $v$-direction for the $\SI{50}{\micro\metre}\times\SI{55}{\micro\metre}$ pixel pitch of the DUT \WfourtyIF\ at perpendicular particle incidence. The hit efficiency averaged over all spatial bins is $99.61\pm0.05\,\si{\percent}$ with the uncertainty given by the standard deviation. The hit efficiency is uniform over all spatial bins, especially also at the borders and corners of the pixels.

\FloatBarrier

\section{Conclusion}
Four production modules of the DEPFET based Belle~II pixel detector, covering all employed pixel pitches, were measured in a high-energy particle beam. All devices under test show low noise and excellent signal-to-noise ratios ranging from 25 to 57. A high hit efficiency of all DUTs, on average \SI{99.6}{\percent}, is demonstrated, even after X-ray irradiation of the silicon oxide up to \SI{266}{\kilo\gray} for one DUT. A hit efficiency of larger than $\SI{99}{\percent}$ is reached for thresholds below \SI{2000}{e^-} for all DUTs. At a threshold of \SI{1400}{e^-}, the sub-pixel efficiency distribution is homogeneous, also in the pixel corners. This paper presents the first measurement of the hit resolution of Belle~II pixel detector modules at perpendicular and non-perpendicular particle incidence. The best spatial resolution is measured to be $5.4\pm1.4\,\si{\micro\metre}$ in $u$-direction for an incidence angle of $\phi=26.9^\circ$ and \SI{50}{\micro\metre} pitch, and $7.0\pm0.6\,\si{\micro\metre}$ ($v$-direction) at $\theta=36.2^\circ$ for a \SI{55}{\micro\metre} pitch. At perpendicular incidence, $11.6\pm0.3\,\si{\micro\metre}$ ($u$-direction) and $13.1\pm0.2\,\si{\micro\metre}$ ($v$-direction, \SI{55}{\micro\metre} pitch) spatial resolution is demonstrated. The measured hit resolutions are compatible with the expectations and allow for reaching the targeted impact parameter resolution for Belle~II of \SI{10}{\micro\metre}~\cite{Kou_2019}. A position finding method that exploits the shape and $\rho$ of a pixel cluster is implemented and yields improvements over the classical centre of gravity method for a large range of incidence angles.  

\section*{Acknowledgements}
We would like to thank the entire Belle~II PXD collaboration for all their hard work that made this measurement possible. The measurements have been performed at the Test Beam Facility at DESY Hamburg (Germany), a member of the Helmholtz Association (HGF). We acknowledge the financial support by the Federal Ministry of Education and Research of Germany and of Generalitat Valenciana (Spain) under the grant number CIDEGENT/2018/020.

\FloatBarrier
\section*{References}

\bibliography{mybibfile}

\end{document}